\UseRawInputEncoding
\documentclass[%
 twocolumn,
nofootinbib,
 amsfonts,amsmath,amssymb,
 aps,
prd,
floatfix,
showpacs,
longbibliography
]{revtex4-2}

\usepackage{graphicx}
\usepackage{dcolumn}
\usepackage{bm}
\usepackage{wasysym}
\usepackage{soul}
\usepackage{xfrac}
\usepackage[dvipsnames]{xcolor}
\definecolor{ReflexBlue}{rgb}{ .0902,.0902,.5882}
\usepackage[breaklinks]{hyperref}
\hypersetup{
    colorlinks=true,
    linkcolor=ReflexBlue,
    filecolor=magenta,      
    urlcolor=ReflexBlue,
    citecolor=ReflexBlue,
}
\usepackage{tikz}
\usetikzlibrary{decorations.pathmorphing}
\usepackage{mathrsfs}

\newcommand{\penrose}{%
\node (ext) at (\y,-\y) {};
\node (int) at (-\y,\y) {};
\path (int)+(90:2*\y)  coordinate (inttop)
           +(-90:2*\y) coordinate (intbot)
           +(0:2*\y)   coordinate (intright)
           +(180:2*\y) coordinate (intleft);
\draw (intleft)--(inttop)--(intright)--(intbot)--(intleft)--cycle;
\path (ext)+(90:2*\y)  coordinate (exttop)
           +(-90:2*\y) coordinate (extbot)
           +(180:2*\y) coordinate (extleft)
           +(0:2*\y)   coordinate (extright);
\draw (extleft)--(exttop)--(extright)--(extbot)--(extleft)--cycle;
}

\def\Tpext{\tikz[baseline=-0.4ex]{
\def\y{4pt}
\penrose
\path (ext)+(\y,\y)   coordinate (a)
           +(-\y,-\y) coordinate (b);
\draw[orange,thick] (a)--(b);
}}
\def\Tmext{\tikz[baseline=-0.4ex]{
\def\y{4pt}
\penrose
\path (ext)+(\y,-\y)     coordinate (a)
           +(-\y,\y) coordinate (b);
\draw[orange,thick] (a)--(b);
}}
\def\Rpext{\tikz[baseline=-0.4ex]{
\def\y{4pt}
\penrose
\path (ext)+(\y,\y)  coordinate (a)
           +(0,0)    coordinate (b)
           +(\y,-\y) coordinate (c);
\draw[orange,thick] (a)--(b)--(c);
}}
\def\Rmext{\tikz[baseline=-0.4ex]{
\def\y{4pt}
\penrose
\path (ext)+(-\y,-\y) coordinate (a)
           +(0,0)     coordinate (b)
           +(-\y,\y)  coordinate (c);
\draw[orange,thick] (a)--(b)--(c);
}}
\def\Tpint{\tikz[baseline=-0.4ex]{
\def\y{4pt}
\penrose
\path (int)+(\y,\y)   coordinate (a)
           +(-\y,-\y) coordinate (b);
\draw[orange,thick] (a)--(b);
}}
\def\Tmint{\tikz[baseline=-0.4ex]{
\def\y{4pt}
\penrose
\path (ext)+(-\y,\y)     coordinate (a)
           +(-3*\y,3*\y) coordinate (b);
\draw[orange,thick] (a)--(b);
}}
\def\Rpint{\tikz[baseline=-0.4ex]{
\def\y{4pt}
\penrose
\path (int)+(\y,-\y)      coordinate (a)
           +(0,0)         coordinate (b)
           +(\y,\y)       coordinate (c);
\draw[orange,thick] (a)--(b)--(c);
}}
\def\Rmint{\tikz[baseline=-0.4ex]{
\def\y{4pt}
\penrose
\path (int)+(-\y,-\y) coordinate (a)
           +(0,0)     coordinate (b)
           +(-\y,\y)  coordinate (c);
\draw[orange,thick] (a)--(b)--(c);
}}
\def\TmextTmint{\tikz[baseline=-0.4ex]{
\def\y{4pt}
\penrose
\path (ext)+(\y,-\y)     coordinate (a)
           +(-3*\y,3*\y) coordinate (b);
\draw[orange,thick] (a)--(b);
}}
\def\TmextRpint{\tikz[baseline=-0.4ex]{
\def\y{4pt}
\penrose
\path (int)+(3*\y,-3*\y)  coordinate (a)
           +(0,0)         coordinate (b)
           +(\y,\y)       coordinate (c);
\draw[orange,thick] (a)--(b)--(c);
}}
\def\RmextTmint{\tikz[baseline=-0.4ex]{
\def\y{4pt}
\penrose
\path (ext)+(-\y,-\y)    coordinate (a)
           +(0,0)        coordinate (b)
           +(-3*\y,3*\y) coordinate (c);
\draw[orange,thick] (a)--(b)--(c);
}}
\def\RmextRpint{\tikz[baseline=-0.4ex]{
\def\y{4pt}
\penrose
\path (ext)+(-\y,-\y)    coordinate (a)
           +(0,0)        coordinate (b)
           +(-2*\y,2*\y) coordinate (c)
           +(-\y,3*\y)   coordinate (d);
\draw[orange,thick] (a)--(b)--(c)--(d);
}}

\begin{document}


\title{Hawking radiation inside a rotating black hole}

\author{Tyler McMaken}
 \email{Tyler.McMaken@colorado.edu}
\author{Andrew J. S. Hamilton}
 \email{Andrew.Hamilton@colorado.edu}
\affiliation{%
 JILA and Department of Physics, University of Colorado, Boulder, Colorado 80309, USA
}%
\date{\today}

\begin{abstract}
In semiclassical gravity, the vacuum expectation value ${\langle\hat{N}\rangle}$ of the particle number operator for a quantum field gives rise to the perception of thermal radiation in the vicinity of a black hole. This Hawking effect has been examined only for observers asymptotically far from a Kerr black hole; here we generalize the analysis to various classes of freely falling observers both outside and inside the Kerr event horizon. Of note, we find that the effective temperature of the ${\langle\hat{N}\rangle}$ distribution remains regular for observers at the event horizon but becomes negative and divergent for observers reaching the (Cauchy) inner horizon. Furthermore, the perception of Hawking radiation varies greatly for different classes of observers, though the spectrum is generally a graybody that decreases in intensity with black hole spin and increases in temperature when looking toward the edges of the black hole shadow.
\end{abstract}

\maketitle

\section{\label{sec:int}Introduction}

If a classical black hole that formed from a gravitational collapse is immersed within a quantum field initially in a vacuum state, someone far away from that black hole will eventually detect excitations of that quantum field in an effect known as Hawking radiation \cite{haw74}. This radiation was found to follow a thermal distribution in the geometric optics (high-frequency) limit, with a temperature proportional to the surface gravity $\varkappa_+$ of the black hole at the event horizon. The key feature required for such radiation to exist is a characteristic exponential redshifting of modes near a (quasi-)trapping horizon. As a result, the Hawking effect can also be related to the radiation seen by, e.g., an accelerating observer or a moving mirror model, where such a redshifting also occurs \cite{bar11a,bar11b}.

The Hawking radiation detected asymptotically far from a black hole is negligibly small for all known astrophysical black holes, orders of magnitude below current observational capabilities. However, the radiation can take on a substantially different form when an observer approaches and/or falls into a black hole. For such an observer, instead of seeing a Hawking temperature proportional to the surface gravity $\varkappa_+$, one can define an effective temperature function ${\kappa}$ that tracks the rate of redshifting they perceive, and this ${\kappa}$ reproduces the thermal Hawking result when a suitable adiabatic condition is met (see Sec.~\ref{subsec:geo} for more details) \cite{bar11a,bar11b}. One may wonder whether the Hawking temperature closer to a black hole's event horizon may be high enough to observe secondary astrophysical effects, but more importantly, the Hawking flux detected \emph{inside} a black hole can be enormous and has profound implications for the self-consistency of black holes models in semiclassical gravity (i.e., quantum field theory placed over a classical background).

The perception of Hawking radiation has been analyzed for various classes of observers during and after gravitational collapse in the Schwarzschild exterior \cite{gre09,bar11c}, the Schwarzschild interior \cite{sai16,ham18}, and most recently, the Reissner-Nordstr\"om exterior and interior \cite{mcm23a}. The goal of the present work is to extend this analysis to the late-time behavior of rotating black holes described by the Kerr metric.

What should one hope to see when analyzing the Hawking content in the Kerr spacetime? Several results may be anticipated from prior studies:

\begin{enumerate}
    \item For an inertial observer in the vicinity of the \emph{event horizon}, the effective temperature has roughly the same order of magnitude as the standard tiny Hawking temperature at infinity (i.e., the event horizon is semiclassically well behaved) \cite{gre09,bar11c,sai16,ham18}.
    \item For an observer in the vicinity of the \emph{inner horizon}, the effective temperature is negative and diverges in the same manner as the Penrose blueshift perturbation singularity \cite{mcm23a,pen68,sim73}.
    \item Hawking radiation is not confined to the radial direction\textemdash an observer looking in an arbitrary direction in their field of view will still see the characteristic exponential redshifting of modes, with higher Hawking temperatures toward the edge of the black hole's shadow and an increasingly isotropic distribution as they approach the inner horizon \cite{ham18,mcm23a}.
    \item The Hawking temperature can become negative even \emph{outside} of the event horizon for a black hole close enough to extremality (e.g.\ for a Reissner-Nordstr\"om charge ${Q/M>\sqrt{8/9}}$, although adiabaticity may not necessarily be satisfied there) \cite{mcm23a}.
\end{enumerate}

One may expect to see similar features for Hawking radiation in the Kerr spacetime, which more closely models astrophysical black holes than simpler, non-rotating models. Such results would confirm through entirely analytical means the same semiclassical divergence recently seen numerically for the renormalized stress-energy tensor at the Kerr Cauchy horizon \cite{zil22b}. This divergence points towards a semiclassical form of the cosmic censorship conjecture, that quantum effects will always act to close off Cauchy horizons that would otherwise serve as entryways to wormholes and timelike singularities. Though the true quantum gravitational nature of a black hole interior remains elusive, these first-order results from quantum field theory over curved spacetime imply either that the Cauchy horizon is the source of a roiling quantum atmosphere that marks the boundary endpoint of spacetime itself, or that the Cauchy horizon is so unstable that it will evaporate outward to meet the event horizon within a matter of seconds to form an extremal black hole or a compact horizonless object \cite{bar21}.

The perception of Hawking radiation for various Kerr observers is explored in Sec.~\ref{sec:eff} using the geometric optics effective temperature formalism, first in the radial direction and then in an arbitrary direction in the observer's field of view. However, these results not only are unreliable at low frequencies, but they also depend crucially on the adiabaticity of the observer at each point of interest. To address both of these concerns, in Sec.~\ref{sec:bog} a full numerical analysis of the wave scattering problem is performed in order to calculate the Bogoliubov spectrum of Hawking radiation in the limits where such a calculation can be feasibly done; in particular, for an observer at infinity, at the event horizon, and at the ingoing and outgoing portions of the Cauchy horizon. For a Reissner-Nordstr\"om black hole, such a calculation led to the conclusion that the Hawking spectrum appears as a graybody at the event horizon but becomes ultraviolet-divergent at the Cauchy horizon, in accord with the geometric optics effective temperature results \cite{mcm23a}. The Kerr spectra computed here follow the same trends as in the spherically symmetric case, except that here we are able to extend the calculations to high enough frequencies to show that the Cauchy horizon radiation does not actually diverge in the ultraviolet regime in most cases (nor should it be expected to\textemdash see Sec.~\ref{subsec:bogs=0} for more details). However, as an observer approaches the Cauchy horizon, they should in general see Hawking radiation glowing brightly in every direction they look, as if they are diving into a thick quantum atmosphere with ever-increasing energy.

\section{\label{sec:eff}Effective temperature for an infaller}

The perception of Hawking radiation in the high-frequency limit is thermal and analytically calculable using an effective temperature function $\kappa(\tau_\text{ob})$. This section explores how this temperature changes for various observers around and inside a Kerr black hole.

In what follows, the mathematical and physical formalism to calculate $\kappa$ is detailed in Secs.~\ref{subsec:set} (\textit{Setup}), \ref{subsec:geo} (\textit{Geometric optics approximation}), and \ref{subsec:unr} (\textit{Unruh state construction}). The effective temperature $\kappa$ is then calculated for two special cases of privileged observers in Sec.~\ref{subsec:pri}: (1) infallers along the black hole's axis of rotation and (2) what we call ``horizostationary'' orbiters. Finally, the temperature for arbitrary freely falling observers, looking in an arbitrary direction, is calculated in Sec.~\ref{subsec:ffo}, with special focus on two cases: (1) equatorial observers with zero angular momentum (ZAMOs), and (2) equatorial observers with zero energy (interior Carter observers).

\subsection{\label{subsec:set}Setup: Kerr metric and quantum field}
Consider a quantum field placed over a fixed Kerr spacetime, which is given by the line element (in Boyer-Lindquist coordinates)\footnote{Throughout this paper we use the ${({-}{+}{+}{+})}$ metric signature and geometric units where ${c=G=k_B=\hbar=1}$.} \cite{car68}
\begin{align}\label{eq:kerr}
    ds^2&=\frac{\rho^2}{\Delta}dr^2-\frac{\Delta}{\rho^2}\left(a\sin^2\!\theta\ d\varphi-dt\right)^2\nonumber\\
    &+\rho^2d\theta^2+\frac{\sin^2\!\theta}{\rho^2}\left(R^2d\varphi-a\ dt\right)^2,
\end{align}
where ${R^2\equiv r^2+a^2}$, where ${a\equiv J/M}$ is the black hole's spin parameter (in terms of the black hole's angular momentum $J$ and mass $M$), the conformal factor ${\rho^2\equiv r^2+a^2\cos^2\!\theta}$ contains zeros at the black hole's ring singularity, and the horizon function ${\Delta\equiv r^2+a^2-2Mr}$ contains zeros at the black hole's event (${r=r_+}$) and Cauchy (${r=r_-}$) horizons.

The geodesic equations of motion in this spacetime are separable \cite{car68}:
\begin{subequations}\label{eq:EOM}
\begin{align}
    \dot{t}&=\frac{1}{\rho^2}\left(\frac{R^2P_r}{\Delta}+aP_\theta\right),\label{eq:EOM_t}\\
    \dot{r}^2&=\frac{1}{\rho^4}\left(P_r^2-\left(K+r^2\delta\right)\Delta\right),\label{eq:EOM_r}\\
    \dot{\theta}^2&=\frac{1}{\rho^4}\left(K-a^2\cos^2\!\theta\ \delta-\frac{P_\theta^2}{\sin^2\!\theta}\right),\label{eq:EOM_theta}\\
    \dot{\varphi}&=\frac{1}{\rho^2}\left(\frac{aP_r}{\Delta}+\frac{P_\theta}{\sin^2\!\theta}\right),
\end{align}
\end{subequations}
where
\begin{subequations}\label{eq:Prth}
\begin{align}
    P_r(r)&\equiv R^2E-aL,\\
    P_\theta(\theta)&\equiv L-aE\sin^2\!\theta,
\end{align}
\end{subequations}
with an overdot representing differentiation with respect to affine time ($\tau$ for massive geodesics and $\lambda$ for massless geodesics), with constants of motion written in terms of the Killing energy per unit mass $E$, Killing angular momentum along the axis of rotation per unit mass $L$, and Carter constant ${K=Q+(aE-L)^2}$, and where ${\delta=1}$ for massive particles while ${\delta=0}$ for massless particles (which will be denoted with scripted constants of motion $\mathcal{E}$, $\mathcal{L}$, $\mathcal{K}$ in contrast to the massive particle's constants $E$, $L$, $K$).

The form of the line element in Eq.~(\ref{eq:kerr}) is unique in that it encodes a special locally inertial, orthogonal frame of reference called the Carter tetrad \cite{car68}. This tetrad $\bm{\gamma}_{\hat{m}}={\{\bm{\gamma}_0,\bm{\gamma}_1,\bm{\gamma}_2,\bm{\gamma}_3\}}$, like any tetrad, is locally flat (${\bm{\gamma}_{\hat{m}}\cdot\bm{\gamma}_{\hat{n}}=\eta_{\hat{m}\hat{n}}}$) and is encoded by the line element via a vierbein $e^{\hat{m}}_{\ \mu}$,
\begin{equation}
    ds^2=e^{\hat{m}}_{\ \mu}e^{\hat{n}}_{\ \nu}\bm{\gamma}_{\hat{m}}\cdot\bm{\gamma}_{\hat{n}} dx^\mu dx^\nu.
\end{equation}
The Carter frame is one of the natural generalizations of the static frame (for Schwarzschild and Reissner-Nordstr\"om black holes) into a stationary frame for a rotating spacetime like Kerr. The Carter frame is particularly special in that it is the only stationary tetrad in which the principal null congruences are purely in the radial direction. The (exterior) Carter vierbein reads \cite{mar83}:
\begin{subequations}\label{eq:carter}
\begin{align}
    e_0^{\ \mu}\partial_\mu&=\frac{R^2\partial_t+a\partial_\varphi}{\rho\sqrt{|\Delta|}},\\
    e_1^{\ \mu}\partial_\mu&=\text{sgn}(\Delta)\frac{\sqrt{|\Delta|}}{\rho}\partial_r,\\
    e_2^{\ \mu}\partial_\mu&=\frac{1}{\rho}\partial_\theta,\\
    e_3^{\ \mu}\partial_\mu&=-\frac{a\sin^2\!\theta\ \partial_t+\partial_\varphi}{\rho\sin\theta}.
\end{align}
\end{subequations}
An observer at rest in this tetrad frame (subsequently referred to as a Carter observer) can exist as a stationary observer anywhere outside of the event horizon. Inside the event horizon, a similar frame can be defined (in particular, swapping ${e_0^{\ \mu}\leftrightarrow e_1^{\ \mu}}$) that hosts an interior Carter observer, who remains ``stationary'' in the spacelike coordinate $t$ (i.e., has zero energy $E$) \cite{mcm21}.

Over the Kerr background one can place a canonically quantized, massless, bosonic field ${{}_s\!\Phi(x)}$ with spin weight\footnote{We follow the notation of Teukolsky \cite{teu72} in using the term ``spin weight'' for the parameter $s$, which either equals the (positive-valued) spin of the field when the ingoing component of the wave is the dominant propagating mode, or the negative of the field's spin when the outgoing component of the wave is the dominant propagating mode. However, $s$ is \emph{not} the same as the spin weight defined in the GHP formalism \cite{ger73} as the eigenvalue of the Lorentz-invariant chiral spin operator and instead would there be called the ``boost weight,'' the eigenvalue of the generator of Lorentz boosts. See Sec.~IIIC of Ref.~\cite{ham22} for more details.\label{foot:spinweight}} $s$. Due to the axial symmetry of the metric encoded by Eq.~(\ref{eq:kerr}), the field ${}_s\!\Phi(x)$ can be decomposed into a complete set of modes ${}_s\phi_{\omega\ell m}(x)$, each accompanied by creation and annihilation operators $a^\dagger$ and $a$,
\begin{equation}\label{eq:modedecomp}
    {}_s\!\Phi(x)=\int_0^\infty d\omega\sum_{\ell=0}^\infty\sum_{m=-\ell}^\ell\left({}_s\phi_{\omega\ell m}a_{\omega\ell m}+{}_s\phi^*_{\omega\ell m}a^\dagger_{\omega\ell m}\right),
\end{equation}
where
\begin{equation}\label{eq:modesep}
    {}_s\phi_{\omega\ell m}=\frac{{}_sf_{\omega\ell}(r,t,\varphi)\ {}_sS^\omega_{\ell m}(\theta)}{R\sqrt{4\pi\omega}}
\end{equation}
(the additional factor of $R$ is included here as in Ref.~\cite{zil22a} so that, among other reasons, the Wronskian of the wave equation will be constant in $r$). Focusing on the scalar (spin-0) case and dropping the spin index in what follows for simplicity, the quantum numbers are the frequency ${\omega\in\mathbb{R}}$, the multipolar number ${\ell\in\mathbb{Z}_{\geq0}}$, and the azimuthal number ${m\in\mathbb{Z}_{\leq\ell}\cap\mathbb{Z}_{\geq-\ell}}$. Thanks to azimuthal and time translation invariance, the mode function ${f_{\omega\ell}(r,t,\varphi)}$ may be further separated as
\begin{equation}\label{eq:modesep_f}
    f_{\omega\ell}(r,t,\varphi)=\psi_{\omega\ell}(r)\ \text{e}^{\pm i\omega t}\ \text{e}^{\pm im\varphi}.
\end{equation}

If the scalar field ${\Phi(x)}$ obeys the Klein-Gordon wave equation ${\Box\Phi=0}$, then the polar function ${S^\omega_{\ell m}(\theta)}$ will satisfy the equation for spheroidal wave functions \cite{ber06}, while the radial function ${\psi_{\omega\ell}(r)}$ will satisfy the radial Teukolsky equation \cite{teu72}
\begin{equation}\label{eq:waveeq_r}
    \frac{d^2\psi_{\omega\ell}}{dr^{*2}}+V_{\omega\ell m}\psi_{\omega\ell}=0,
\end{equation}
with the scattering potential
\begin{equation}\label{eq:V}
    V_{\omega\ell m}\equiv\left(\omega-\frac{ma}{R^2}\right)^2-\frac{\lambda_{\omega\ell m}\tilde{\Delta}}{r}-\tilde{\Delta}^2-\frac{d\tilde{\Delta}}{dr^*}.
\end{equation}
In Eq.~(\ref{eq:V}), the constant $\lambda_{\omega\ell m}$ is defined in Appendix~\ref{app:mst} below Eq.~(\ref{eq:waveeq_r_R}), the tortoise coordinate $r^*$ is defined by
\begin{equation}\label{eq:tort}
    \frac{dr}{dr^*}=\frac{\Delta}{R^2},
\end{equation}
and the function $\tilde{\Delta}$ is defined by
\begin{equation}
    \tilde{\Delta}(r)\equiv\frac{r\Delta}{R^4}.
\end{equation}

\subsection{\label{subsec:geo}Geometric optics approximation}
In the eikonal (geometric optics, or high-frequency) approximation, the above field's wave equation is solved with the Ansatz
\begin{equation}
    \phi_{\omega\ell m}(x)=\mathcal{A}(x)\ \text{e}^{i\omega\Theta(x)},
\end{equation}
which leads to an equation for the eikonal phase function $\Theta(x)$ at leading order in inverse powers of $\omega$,
\begin{equation}\label{eq:eikonal}
    \partial^\mu\Theta\partial_\mu\Theta=0.
\end{equation}
It can be shown by covariant differentiation of Eq.~(\ref{eq:eikonal}) that it is a geodesic equation for a null vector field ${k^\mu\equiv\partial^\mu\Theta}$ normal to the family of constant-$\Theta$ hypersurfaces. Thus, any wave scattering problem can be reduced in the geometric optics limit to a ray-tracing problem along the eikonal hypersurface-orthogonal null congruence.

The scattering problem under question is the problem of finding the Bogoliubov coefficient between the vacuum state of an observer and the Unruh vacuum state in the asymptotic past. That is, if the annihilation operators of Eq.~(\ref{eq:modedecomp}) define an observer's vacuum state ${|0_\text{ob}\rangle}$ via
\begin{equation}
    a_\text{ob}|0_\text{ob}\rangle=0
\end{equation}
(suppressing quantum number indices), and a completely equivalent decomposition into a set of modes $\bar{\omega}$, $\bar{\ell}$, and $\bar{m}$ defines the vacuum state of an emitter in the past of the observer via
\begin{equation}
    a_\text{em}|0_\text{em}\rangle=0,
\end{equation}
then the Bogoliubov spectrum of Hawking radiation will be given by the expectation value of the observer's number operator in the emitter's vacuum:
\begin{equation}\label{eq:bog}
    \langle0_\text{em}|a^{\dagger}_\text{ob}a_\text{ob}|0_\text{em}\rangle=\int_0^\infty d\bar{\omega}\sum_{\bar{\ell}=0}^\infty\sum_{\bar{m}=-\bar{\ell}}^{\bar{\ell}}\left|\langle\phi_\text{em}|\phi^*_\text{ob}\rangle\right|^2,
\end{equation}
where bra-ket notation denotes the Lorentz-invariant Klein-Gordon inner product, which consists of a 3D integral over an arbitrary spacelike Cauchy hypersurface $\Sigma$ that terminates at spacelike infinity and is orthogonal to a future-directed unit vector $n^\mu$:
\begin{equation}\label{eq:inner_product}
    \langle\phi_1|\phi_2\rangle\equiv-i\int_\Sigma d\Sigma\ n^\mu\sqrt{-g_\Sigma}\ \phi_1\overset{\leftrightarrow}{\partial}_\mu\phi_2^*,
\end{equation}
where the bidirectional derivative $\overset{\leftrightarrow}{\partial}_\mu$ is defined below Eq.~(\ref{eq:scalarproductintegrals}).

In the eikonal approximation, Eq.~(\ref{eq:bog}) will yield a blackbody spectrum with a temperature ${\kappa/(2\pi)}$, as long as $\kappa$ is defined as the rate of exponential redshift for a null ray connecting a vacuum-state emitter to an infalling observer \cite{bar11a,bar11b,ham18}:
\begin{equation}\label{eq:kappa_tau}
    \kappa(\tau_\text{ob})=-\frac{d}{d\tau_\text{ob}}\ln\left(\frac{\omega_\text{ob}}{\omega_\text{em}}\right),
\end{equation}
where ${\omega_i\equiv-k^\mu\dot{x}_\mu}$ is the temporal component of the 4-velocity of a null particle measured in the frame of the observer ($i=$``ob'') or the emitter ($i=$``em''), so that the proper time derivative of this frequency $\omega_i$ measures the null particle's redshift. This classical frequency $\omega_i$ will always be presented with a subscript to distinguish it from the frequency $\omega$ of the quantum mode $\phi_{\omega\ell m}$ in Eq.~(\ref{eq:modedecomp}) and following.

The function $\kappa$ defined by Eq.~(\ref{eq:kappa_tau}) is here called the effective temperature, since it reproduces the Hawking temperature in the geometric optics limit, so long as the following adiabatic condition is met \cite{bar11c}:
\begin{equation}\label{eq:epsilon}
    \epsilon(\tau_\text{ob})\equiv\frac{1}{\kappa^2}\left|\frac{d\kappa}{d\tau_\text{ob}}\right|\ll1.
\end{equation}
It should be noted that even when the adiabatic control function $\epsilon$ defined by Eq.~(\ref{eq:epsilon}) is not small, a non-zero effective temperature $\kappa$ should still generally imply the presence of Hawking particles in the frame of the observer, though those particles may not necessarily follow a thermal spectrum \cite{bar11a}.

\subsection{\label{subsec:unr}Unruh state construction}

In the effective temperature formalism, the appropriate choice of vacuum state for gravitational collapse leading to the formation of a black hole is that of an inertial emitter in the asymptotic past, when the spacetime is still flat and the black hole has not yet formed. However, the Kerr metric models an eternal black hole (or, by analytic extension, a white hole-black hole system), not a dynamical collapse. Thus, instead of starting with a Minkowski vacuum in the asymptotic past, one must specify boundary conditions on the Kerr past horizon that match the exponential redshifting one would expect near a collapsing shell of matter, and these boundary conditions are precisely the ones used to define the (past) Unruh vacuum state used here. This state exactly mimics the physical state coming from gravitational collapse in all portions of the spacetime except along the (now singular) past horizon and along the left leg of the inner horizon (which in the dynamical case may not be a Cauchy horizon as it is in Kerr; see Ref.~\cite{kra06} for a proposed construction in the analogous charged case).

The Unruh state is formally defined by taking modes to be positive frequency with respect to the Killing vector field ${\partial_t}$ along past null infinity and with respect to the Kruskalized canonical affine field ${\partial_U}$ along the past horizon \cite{unr76}. The latter coordinate is defined in the physical regions of interest by
\begin{equation}\label{eq:U}
    U\equiv\frac{\text{sgn}(r_+-r)}{\varkappa_+}\ \text{e}^{-\varkappa_+u},
\end{equation}
where
\begin{equation}\label{eq:surf}
    \varkappa_\pm\equiv\frac{1}{2R_\pm^2}\frac{d\Delta}{dr}\bigg|_{r_\pm}=\pm\frac{r_+-r_-}{2R_\pm^2}
\end{equation}
is the surface gravity at the black hole's event horizon ($\varkappa_+$) or Cauchy horizon ($\varkappa_-$) with the definition ${R_\pm^2\equiv r_\pm^2+a^2}$, and
\begin{equation}\label{eq:uv}
    u\equiv t-r^*,\qquad v\equiv t+r^*
\end{equation}
are the outgoing and ingoing Eddington-Finkelstein coordinates, defined in the same way for both the interior and exterior portions of the spacetime.

Since the definition of a vacuum state primarily concerns the choice of positive frequency with respect to a timelike coordinate, the choice of angular modes will not substantially influence the final results of the calculations done here \cite{zil22a}. For the Unruh modes along past null infinity, azimuthal modes of the form ${\text{exp}(i\bar{m}\varphi)}$ are used, while the Unruh modes along the past horizon are taken to be ${\text{exp}(i\bar{m}\varphi_+)}$, where
\begin{equation}\label{eq:phi_pm}
    \varphi_\pm\equiv\varphi-\Omega_\pm t,
\end{equation}
with the angular velocity $\Omega_\pm$ of the horizon at ${r=r_\pm}$ defined in Eqs.~(\ref{eq:Omega_p}) and (\ref{eq:omega_pm}). The azimuthal coordinate $\varphi_+$ is regular at the horizon, and additionally, it defines the Killing vector ${\partial_t+\Omega_+\partial_\varphi}$ that generates the Killing horizon at ${r=r_+}$.

In what follows, it will be shown that the Unruh state can be encoded in the geometric optics framework by a family of phase-aligned, freely falling emitters placed at ${r\to\infty}$ for ingoing modes and ${r\to r_+}$ for outgoing modes. Consider first the ingoing Unruh sector, which is defined with no mode contributions from the past horizon and with modes of the form ${\text{exp}(-i\bar{\omega}v)}$ along past null infinity \cite{unr76}. In the geometric optics limit, these ingoing modes should follow a null congruence hypersurface-orthogonal to the eikonal phase front defined by taking ${\Theta=v}$ along past null infinity. However, these null geodesics could just as easily be labeled by the proper time of an infaller asymptotically far from the black hole: from Eqs.~(\ref{eq:EOM}), (\ref{eq:uv}), and (\ref{eq:tort}), as an infaller's radius $r$ is taken to infinity, the ingoing time behaves as
\begin{equation}\label{eq:vdot_rinf}
    \lim_{r\to\infty}\frac{dv}{d\tau}=\lim_{r\to\infty}\left(\dot{t}+\frac{R^2}{\Delta}\dot{r}\right)=E-\sqrt{E^2-1}.
\end{equation}
If the infaller is taken to be at rest asymptotically far from the black hole (${E=1}$), then Eq.~(\ref{eq:vdot_rinf}) implies that ${d\tau=dv}$; i.e., the infaller's proper time will tick at the same rate as the null coordinate used to define the Unruh state at past null infinity.

Now consider the outgoing Unruh sector, which is defined with no mode contributions from past null infinity and with modes of the form ${\text{exp}(-i\bar{\omega}U)}$ along the past horizon \cite{unr76}. At the past horizon, when ${r\to r_+}$ (and ${\Delta\to0}$), the rate of change of an infaller's outgoing Eddington-Finkelstein coordinate $u$ with respect to their proper time will diverge:
\begin{equation}\label{eq:udot_r+}
    \lim_{r\to r_+}\frac{du}{d\tau}=\frac{2R^2P_r}{\rho^2\Delta}+\mathcal{O}(1).
\end{equation}

In order to show that the appropriate choice of coordinate is actually $U$ instead of $u$, consider how $u$ explicitly depends on an infaller's proper time. First, define the Mino time ${\tilde{\tau}}$ \cite{min03} by the relation
\begin{equation}\label{eq:mino}
    \frac{d\tau}{d\tilde{\tau}}=\rho^2
\end{equation}
so that as ${r\to r_+}$, Eq.~(\ref{eq:EOM_r}) can be integrated to yield the asymptotic timelike geodesic solution
\begin{equation}\label{eq:tau_rr+}
    \lim_{r\to r_+}\tilde{\tau}=
    \tilde{\tau}_0-\frac{r-r_+}{|P_r|}+\mathcal{O}\left[(r-r_+)^2\right],
\end{equation}
with an integration constant $\tilde{\tau}_0$. Similarly, Eq.~(\ref{eq:EOM_t}) can be integrated in the same asymptotic limit to yield the timelike geodesic solution
\begin{align}\label{eq:t_rr+}
    \lim_{r\to r_+}t&=\lim_{r\to r_+}\left(\int dr\frac{R^2P_r}{(dr/d\tilde{\tau})\Delta}+\int d\theta\frac{aP_\theta}{d\theta/d\tilde{\tau}}\right)\nonumber\\
    &=\int dr\frac{R^2}{\Delta}+\mathcal{O}(1)\nonumber\\
    &=r^*+\mathcal{O}(1),
\end{align}
where the final ${\mathcal{O}(1)}$ term encompasses terms at least constant in $r$, including terms dependent on the latitude $\theta$. The outgoing Eddington-Finkelstein coordinate $u$ can therefore be written from Eqs.~(\ref{eq:uv}) and (\ref{eq:t_rr+}) as
\begin{equation}\label{eq:u_rr+}
    \lim_{r\to r_+}u=-2r^*+\mathcal{O}(1)=-\frac{1}{\varkappa_+}\ln|r-r_+|+\mathcal{O}(1),
\end{equation}
with the surface gravity $\varkappa_+$ from Eq.~(\ref{eq:surf}). Inverting Eq.~(\ref{eq:u_rr+}) and substituting in the inverse of Eq.~(\ref{eq:tau_rr+}) gives the well-known exponential relation
\begin{equation}
    \lim_{r\to r_+}\tilde{\tau}\propto\text{e}^{-\varkappa_+ u}+\mathcal{O}(1),
\end{equation}
which is precisely the relation used to define the Kruskalized coordinate $U$ in Eq.~(\ref{eq:U}). Thus, the proper time of an ingoing infaller asymptotically close to the event horizon labels outgoing null geodesics in the same fashion as the Kruskalized coordinate $U$ used to define the Unruh state.

As can be seen from the analysis above, the choice of the infaller emitting null rays to define the Unruh state is independent of that infaller's orbital parameters and angular position, as long as they begin at rest asymptotically far from the black hole, follow along an ingoing timelike geodesic, and reside either at ${r\to\infty}$ (for ingoing modes) or ${r\to r_+}$ (for outgoing modes).

When considering the geometric optics Unruh state in the Kerr geometry, an additional subtlety arises that is not present in the Schwarzschild or Reissner-Nordstr\"om geometries. For those simpler, spherically symmetric cases, an observer in radial free-fall looking down at an Unruh emitter asymptotically close to the event horizon is able to watch the same emitter for their entire journey into the black hole. However, for a Kerr black hole, an observer in free-fall generally (except for a few privileged frames analyzed in Sec.~\ref{subsec:pri}) cannot watch the same emitter at the horizon without rotating their field of view or otherwise accelerating. The reason for the complication is that the emitter for the outgoing Unruh state is within the ergosphere and must orbit the black hole with the geometry. An observer would therefore not see the redshifting emitter freeze in place as they approach the horizon, but instead at late times they would see the emitter steadily moving across the surface of the past horizon until reaching the edge of the black hole's shadow, becoming heavily distorted, and reappearing on the opposite side.

\begin{figure}[tb]
\centering
\includegraphics[width=0.75\columnwidth]{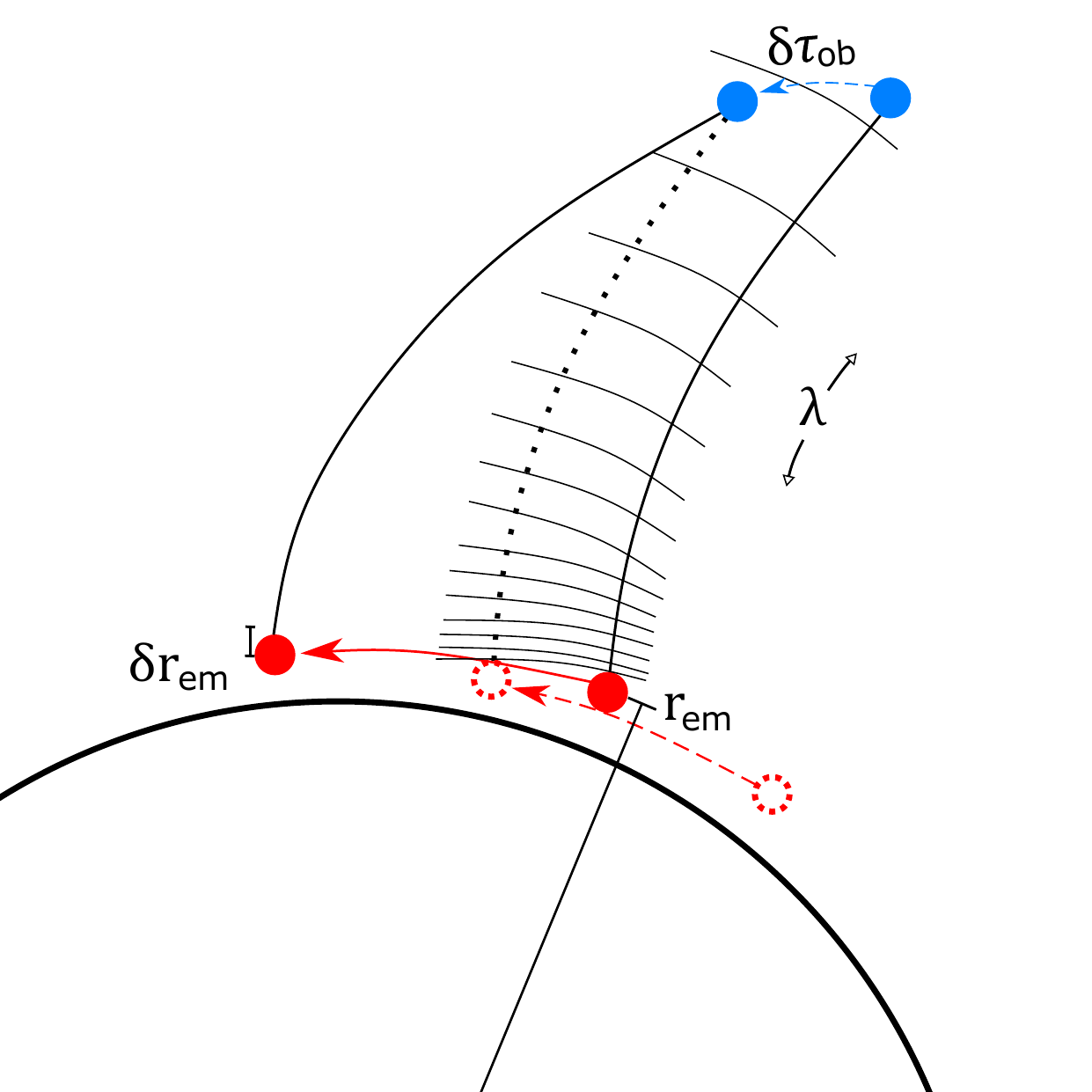}
\caption{An inertial observer (blue) cannot follow outgoing null geodesics from one emitter (solid red) without rotating their frame of reference. But if they stare in a fixed direction, the new emitter they see (dashed red) after an infinitesimal proper time $\delta\tau_\text{ob}$ must be shifted by a radial distance $\delta r_\text{em}$ so that all emitters remain in phase. Then, the total affine distance $\lambda$ will change, but the affine distance weighted by the emitter's frequency $\omega_\text{em}$, Eq.~(\ref{eq:lambda_em_ob}), will stay constant.\label{fig:kerr_wavefronts}}
\end{figure}

As has been argued in previous studies \cite{ham18}, an observer who rotates their frame of reference to follow a single emitter will induce undesired non-inertial particle creation effects, which are fundamentally distinct from the particle creation due to the Hawking effect. For the Kerr geometry, one must therefore consider a \emph{family} of Unruh emitters at the event horizon, all chosen to lie along the same eikonal wave front, so that as the observer falls toward the black hole, their non-rotating (Fermi-Walker transported) view will sweep across different emitters all remaining in phase with each other (see Fig.~\ref{fig:kerr_wavefronts}).

The implementation of a family of phase-aligned Unruh emitters is carried out in Sec.~\ref{subsec:ffo}. The key constraint imposed on the calculation of the effective temperature of Eq.~(\ref{eq:kappa_tau}) is that the affine distance of the null geodesic measured in the frame of the emitter must be held constant for fixed observer position as the emitter's position is varied along the horizon. The affine distance $\lambda$ along a Kerr null geodesic, analogous to the proper time $\tau$ for timelike geodesics, can be obtained by quadrature of Eqs.~(\ref{eq:EOM_r}) and (\ref{eq:EOM_theta}): 
\begin{equation}\label{eq:lambda}
    \lambda=\int_{r_\text{em}}^{r_\text{ob}}\frac{r^2\ dr}{\pm\sqrt{P_r^2-\mathcal{K}\Delta}}+\int_{\theta_\text{em}}^{\theta_\text{ob}}\frac{a^2\cos^2\!\theta\ d\theta}{\pm\sqrt{\mathcal{K}-P_\theta^2\csc^2\!\theta}}.
\end{equation}
This affine distance can be scaled by the null particle's frequency $\omega_i$ to yield proper distances in the frames of the emitter ($\lambda_\text{em}$) or the observer ($\lambda_\text{ob}$):
\begin{equation}\label{eq:lambda_em_ob}
    \lambda=\frac{\lambda_\text{em}}{\omega_\text{em}}=\frac{\lambda_\text{ob}}{\omega_\text{ob}}.
\end{equation}
Eq.~(\ref{eq:lambda_em_ob}) can then be substituted into Eq.~(\ref{eq:kappa_tau}) to give a new expression for the effective temperature $\kappa$, with the constraint that $\lambda_\text{em}$ be kept constant:
\begin{equation}\label{eq:kappa_lambda_em}
    \kappa(\tau_\text{ob})=-\frac{d}{d\tau_\text{ob}}\ln\left(\frac{\omega_\text{ob}\lambda}{\lambda_\text{em}}\right)=-\frac{d\ln\omega_\text{ob}}{d\tau_\text{ob}}-\frac{d\ln\lambda}{d\tau_\text{ob}}.
\end{equation}

\subsection{\label{subsec:pri}Privileged observers}

Before utilizing the constant-phase constraint necessary for general observers sweeping across a family of horizon-limit emitters, consider the special cases where an observer is able to stare at a single emitter throughout their entire free-fall descent. Such privileged frames give rise to feasible analytic calculations of the effective temperature, and while they are usually non-inertial and require an observer to accelerate radially or azimuthally, two exceptional cases will be considered here: on-axis observers free-falling along the ${\theta=0}$ pole, and ``horizostationary'' observers orbiting the black hole at the same angular speed as the event horizon.

\subsubsection{\label{subsubsec:on-axis}On-axis observers}

\begin{figure*}[t]
\centering
\begin{minipage}[l]{0.85\columnwidth}
  \includegraphics[width=\columnwidth]{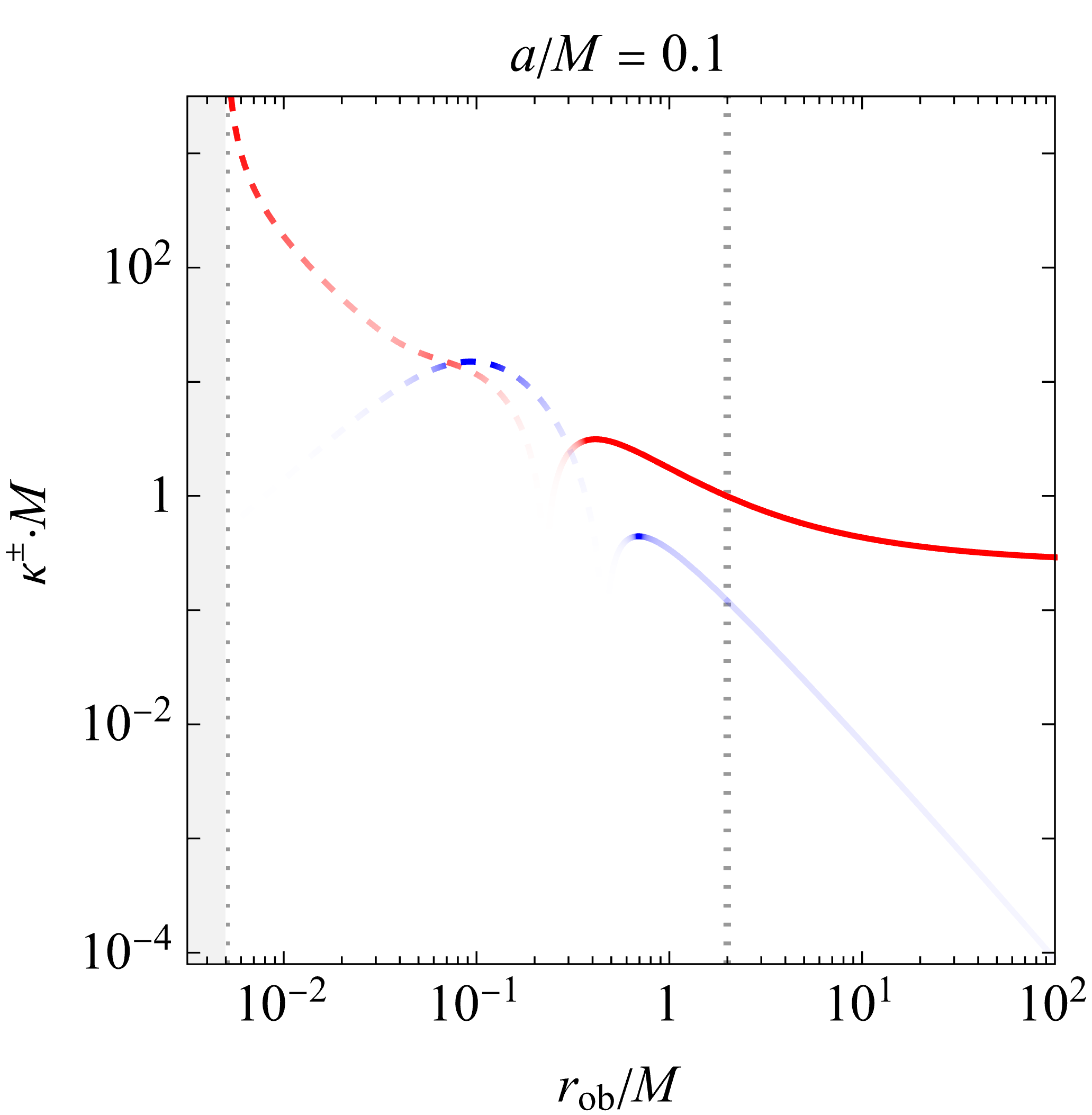}
\end{minipage}%
\hspace{1.5em}
\begin{minipage}[r]{1.02\columnwidth}
    \includegraphics[width=\columnwidth]{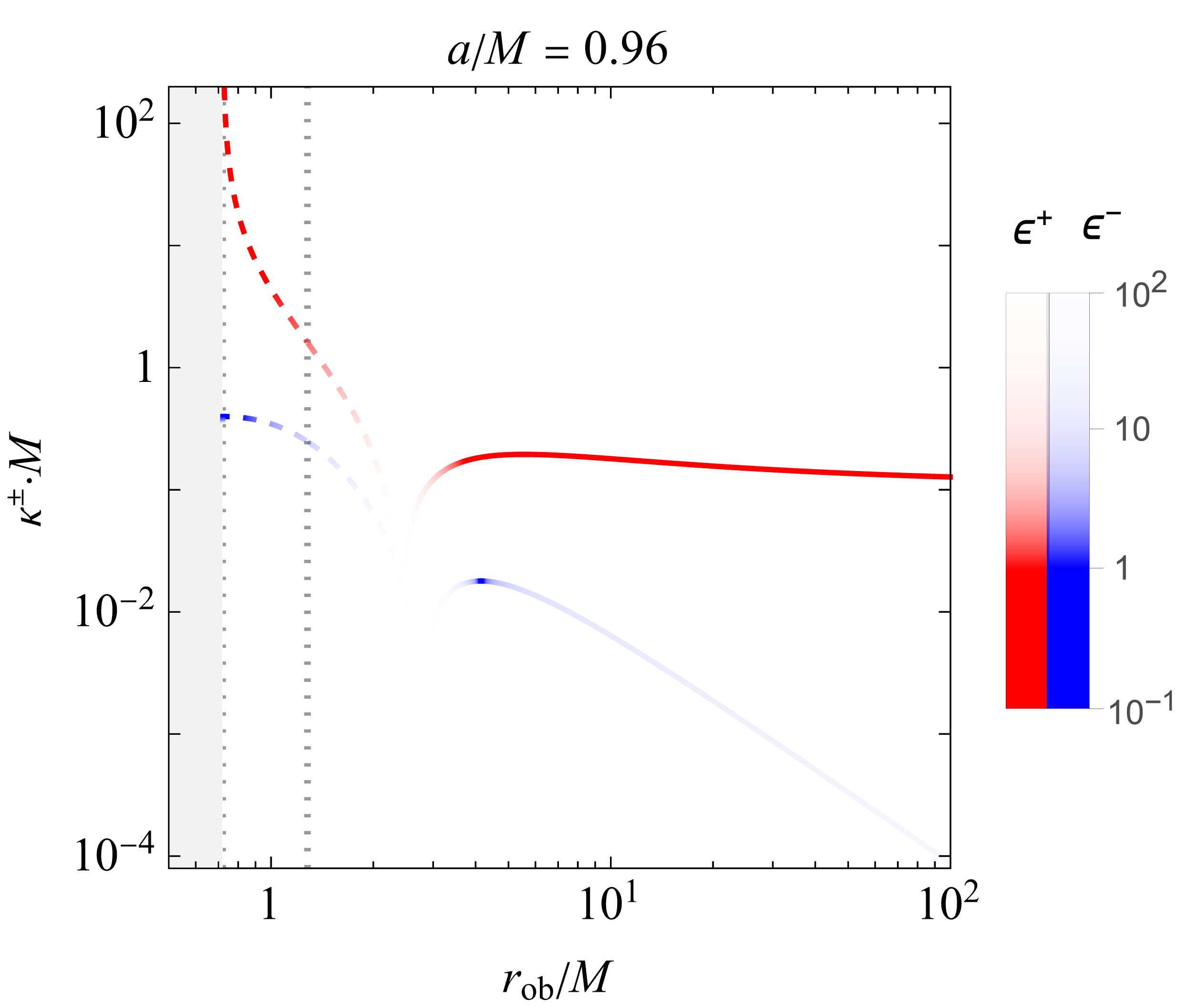}
\end{minipage}
\caption{Effective Hawking temperatures $\kappa^\pm$ seen by an observer freely falling along the Kerr ${\theta=0}$ rotational axis looking directly inward ($\kappa^+$, red curve) or outward ($\kappa^-$, blue curve) at different radii $r_\text{ob}$, for two choices of the Kerr black hole spin parameter $a$, all in units of the black hole mass $M$. Solid curves indicate positive values on the log plot, and dashed curves indicate negative values. Lighter colors indicate higher values of the adiabatic control function $\epsilon^\pm$ from Eq.~(\ref{eq:epsilon}), which imply less confidence in the validity of the geometric optics approximation. The inner and outer horizons are shown with gray, dotted vertical lines, and the unphysical region below the inner horizon is grayed out. When the observer is asymptotically far away, the effective Hawking temperature $\kappa^+$ approaches a constant equal to the surface gravity, but as the observer reaches the inner horizon, the effective temperature becomes negative and diverges.\label{fig:kappa_on-axis}}
\end{figure*}

Geodesics along the rotational axis of a Kerr black hole, where ${\theta=0}$ and ${\dot{\theta}=0}$, must have constants of motion ${L=0}$ and ${K=a^2}$, leading to the geodesic equations
\begin{subequations}\label{eq:EOM_on-axis}
\begin{align}
    \dot{t}&=\frac{R^2}{\Delta}E,\\
    \dot{r}^2&=E^2-\frac{\Delta}{R^2}\left(\frac{r^2\delta+a^2}{R^2}\right),\\
    \dot{\theta}^2&=0,\\
    \dot{\varphi}&=\left(\frac{1}{\Delta}-\frac{1}{R^2}\right)aE,
\end{align}
\end{subequations}
with the same notation as in Eqs.~(\ref{eq:EOM}). A null particle traveling outward ($+$) or inward ($-$) along this axis will be detected by a freely falling observer ($i=$``ob'') or emitter ($i=$``em'') with frequency
\begin{align}
    \omega_i&=-k^\mu\dot{x}_\mu\nonumber\\
    &=\frac{R^2}{\Delta}\left(E\pm\sqrt{\left(E^2-\frac{\Delta}{R^2}\right)\left(1-\frac{a^2\Delta}{R^4}\right)}\right),
\end{align}
normalized to the frequency seen by someone at rest asymptotically far away. This frequency is independent of the rate $\dot{\varphi}$ at which the infaller is rotating with the geometry as they make their descent.

A freely falling observer on the symmetry axis will then detect two independent effective temperatures: if they look directly downward into the pole of the black hole they will see outgoing Hawking modes redshifting from an emitter near the past horizon, and if they look directly upward into the sky they will see ingoing Hawking modes redshifting from an emitter near past null infinity. The calculation of these effective temperatures then proceeds from an application of the chain rule to Eq.~(\ref{eq:kappa_tau}):
\begin{align}\label{eq:kappa_chain_on-axis}
    \kappa^\pm=-\omega_\text{ob}\left(\frac{\dot{r}_\text{ob}}{\omega_\text{ob}}\frac{d\ln\omega_\text{ob}}{dr_\text{ob}}-\frac{\dot{r}_\text{em}}{\omega_\text{em}}\frac{d\ln \omega_\text{em}}{dr_\text{em}}\right),
\end{align}
which makes use of the relation
\begin{equation}
    \frac{d\tau_\text{em}}{d\tau_\text{ob}}=\frac{\omega_\text{ob}}{\omega_\text{em}}.
\end{equation}
An intermediate result is
\begin{align}\label{eq:dlnomegadr_on-axis}
    &\frac{d\ln\omega_i}{dr}=\nonumber\\
    &\frac{\frac{2r}{R^2}\left(1-2\omega_iE+\frac{a^2\Delta}{R^4}\right)-\frac{\Delta'}{\Delta}\left(1-2\omega_iE+\frac{a^2E^2}{R^2}\right)}{2\left(1-\omega_iE+\frac{a^2}{R^2}\left(E^2-\frac{\Delta}{R^2}\right)\right)},
\end{align}
where a prime denotes differentiation with respect to the Boyer-Lindquist radial coordinate $r$.

The effective temperature's dependence on the observer's position $r$ and energy $E$ can then be calculated with the help of Eq.~(\ref{eq:dlnomegadr_on-axis}). The part of Eq.~(\ref{eq:kappa_chain_on-axis}) that depends on the emitter reduces to
\begin{equation}
    \lim_{r_\text{em}\to\infty}\frac{\dot{r}_\text{em}}{\omega_\text{em}}\frac{d\ln \omega_\text{em}}{dr_\text{em}}=0
\end{equation}
for ingoing modes originating from an Unruh emitter asymptotically far from the black hole (i.e., an observer looking straight up at the sky, measuring an effective temperature $\kappa^-$), while it reduces to
\begin{equation}
    \lim_{r_\text{em}\to r_+}\frac{\dot{r}_\text{em}}{\omega_\text{em}}\frac{d\ln \omega_\text{em}}{dr_\text{em}}=\varkappa_+
\end{equation}
for outgoing modes originating from an Unruh emitter asymptotically close to the past horizon (i.e., an observer looking straight down at the black hole, measuring an effective temperature $\kappa^+$).

For an infalling observer with unit energy (${E=1}$) descending along the rotational axis, the effective temperatures seen above and below are shown in Fig.~\ref{fig:kappa_on-axis} for a slowly spinning black hole (${a/M=0.1}$) and a near-extremal one (${a/M=0.96}$). Some analytic limits are worth mentioning explicitly:
\begin{subequations}\label{eq:kappa_asymp_on-axis}
\begin{align}
    \kappa^+&=
    \begin{cases}
        \varkappa_+,&r_\text{ob}\to\infty\\
        \displaystyle\frac{8r_+R_+^2-a^2\Delta'(r_+)}{2R_+^4}-\frac{2}{\Delta'(r_+)},&r_\text{ob}\to r_+\\
        \displaystyle-\frac{R_+^2+R_-^2}{R_+^2}\frac{1}{r-r_-}+\mathcal{O}(1),&r_\text{ob}\to r_-
    \end{cases},\\
    \kappa^-&=
    \begin{cases}
        0,&r_\text{ob}\to\infty\\
        \displaystyle\frac{\varkappa_\pm r_\pm^4-4a^2r_\pm}{2R_\pm^2(r_\pm^2+2a^2)},&r_\text{ob}\to r_\pm
    \end{cases}.
\end{align}
\end{subequations}
From Eqs.~(\ref{eq:kappa_asymp_on-axis}), one can see that an on-axis infaller looking downward will see a Hawking temperature proportional to the surface gravity $\varkappa_+$ from Eq.~(\ref{eq:surf}) when they are asymptotically far away, as expected. However, this effective temperature will change as they approach the black hole\textemdash as they cross the event horizon, the effective temperature will generally increase by a factor of two or so, depending on the black hole's spin $a$ (in accordance with assertion $\#$1 in Sec.~\ref{sec:int}). But just as was seen for an electrically charged black hole \cite{mcm23a}, an event horizon-crossing observer will see a negative temperature ${\kappa^+(r_+)<0}$ when the black hole is close enough to extremality (in accordance with assertion $\#$4 in Sec.~\ref{sec:int}). In the Reissner-Nordstr\"om case, a radial free-faller will see a negative temperature outside the event horizon when ${Q/M>\sqrt{8/9}\sim0.943}$ \cite{mcm23a}, but in the Kerr case, an on-axis free-faller will see a negative temperature outside the event horizon when ${a/M\gtrsim0.860}$. This limiting value is similar but not equal to the spin ${a/M=\sqrt{3}/2}$ at which tidal forces change sign for an on-axis observer crossing the Kerr event horizon \cite{lim20}. However, there is no reason \emph{a priori} why the tidal forces and effective temperatures should agree (though they do in the Reissner-Nordstr\"om case)\textemdash the tidal forces are calculated from the locally measured Riemann curvature tensor and the geodesic deviation equation, which uses a different expansion order compared to the eikonal Eq.~(\ref{eq:eikonal}) used to calculate the effective temperature.

Once the on-axis infaller dips below the event horizon, the effective temperature $\kappa^+$ decreases until, as shown in Eq.~(\ref{eq:kappa_asymp_on-axis}) and Fig.~\ref{fig:kappa_on-axis}, the effective temperature diverges to negative infinity (in accordance with assertion $\#$2 in Sec.~\ref{sec:int}). Such a divergence will always occur at the inner horizon of a stationary, rotating black hole in the Unruh state, as has been shown explicitly in Ref.~\cite{mcm23b}. Even if the observer turns around inside the black hole and acquires ${E<0}$, the outgoing temperature $\kappa^+$ will become finite, but the ingoing temperature $\kappa^-$ will then diverge to negative infinity. The inner horizon is thus the surface beyond which the semiclassical approximation can absolutely no longer be trusted.

\subsubsection{\label{subsubsec:horizo}Horizostationary observers}

The second class of privileged observers to be analyzed are those observers who orbit the black hole with the same angular velocity as an infaller at the event horizon. Much like a satellite in a geostationary orbit above Earth, these ``horizostationary'' observers will hover above the same spot on the event horizon as the black hole rotates, so that they can track the same null ray originating from an Unruh emitter as they travel along their own worldline.

Focusing on a stationary observer orbiting a Kerr black hole in the equatorial plane (${\theta=90^\circ}$), it is known that such an observer will only be freely falling (4-acceleration ${D\dot{x}^\mu/d\tau=0}$) if their angular velocity is \cite{sem93}
\begin{equation}\label{eq:Omega_ffo}
    \Omega\equiv\frac{d\varphi}{dt}=\frac{\sqrt{M}}{a\sqrt{M}\pm r^{3/2}}.
\end{equation}
Geodesic horizostationary observers can thus exist only at a single spin-dependent radius, found by matching Eq.~(\ref{eq:Omega_ffo}) with the angular velocity of an Unruh emitter at the event horizon, which can be found from Eqs.~(\ref{eq:EOM}) to equal
\begin{equation}\label{eq:Omega_p}
    \Omega_+\equiv\frac{a}{R_+^2}.
\end{equation}
While observers can orbit with this angular velocity at any radius above the event horizon, most will be forced to accelerate radially unless they are at the radius
\begin{equation}
    r_\text{HO}=\left(\frac{r_+^4M}{a^2}\right)^{1/3}.
\end{equation}
This radius tends to infinity in the Schwarzschild ${a=0}$ limit and to the event horizon in the extremal ${a=M}$ limit.

A horizostationary observer at radius $r_\text{HO}$ in the equatorial plane will have geodesic equations of motion
\begin{subequations}\label{eq:EOM_HO}
\begin{align}
    \dot{t}&=\frac{1}{r^2}\left(\frac{R^2P_r}{\Delta}+aP_\theta\right),\\
    \dot{r}^2&=0,\\
    \dot{\theta}^2&=0,\\
    \dot{\varphi}&=\frac{1}{r^2}\left(\frac{aP_r}{\Delta}+P_\theta\right),
\end{align}
\end{subequations}
where the constants $P_r$ and $P_\theta$, defined by Eqs.~(\ref{eq:Prth}) evaluated at ${r=r_\text{HO}}$ and ${\theta=90^\circ}$, are fixed by the constraints ${\Omega_+=\dot{\varphi}/\dot{t}}$ and ${P_r^2=(P_\theta^2+r^2)\Delta}$.

Since the horizostationary observer only moves along the Kerr metric's Killing fields ${\partial_t}$ and ${\partial_\varphi}$, the frequency $\omega_\text{ob}$ of an outgoing null geodesic seen by the observer will not change with the observer's proper time $\tau_\text{ob}$. Thus, from Eq.~(\ref{eq:kappa_tau}), the only dynamic contribution to the outgoing effective temperature $\kappa^+$ will be from the freely falling emitter at the event horizon:
\begin{equation}\label{eq:kappa_chain_HO}
    \kappa^+=\frac{\omega_\text{ob}}{\omega_\text{em}}\frac{d\ln\omega_\text{em}}{d\tau_\text{em}}.
\end{equation}

The frequency $\omega_\text{em}$ of an outgoing equatorial null particle with dimensionless orbital parameters $\mathcal{L}/\mathcal{E}$ and $\mathcal{K}/\mathcal{E}^2=(a-\mathcal{L}/\mathcal{E})^2$, measured in the frame of an infalling equatorial emitter with constants of motion $E$, $L$, and ${K=(aE-L)^2}$, is
\begin{align}\label{eq:omega_em_HO}
    &\omega_\text{em}=\frac{1}{\Delta}\Bigg(R^2E-\mathcal{L}L+\frac{2}{r}(aE-L)(a-\mathcal{L})\nonumber\\
    &+\sqrt{\left(R^2-\mathcal{L}^2+\frac{2}{r}(a-\mathcal{L})^2\right)\left(\frac{P_r^2}{r^2}-\left(1+\frac{P_\theta^2}{r^2}\right)\Delta\right)}\Bigg),
\end{align}
where the photon energy $\mathcal{E}$ is set to unity without loss of generality. Even though this frequency $\omega_\text{em}$ depends on the emitter's orbital parameters via $E$, $L$, $P_r$, and $P_\theta$, the effective temperature $\kappa^+$ will be independent of the emitter's motion once the emitter is taken to be asymptotically close to the event horizon, as argued in Sec.~\ref{subsec:unr}. If the Unruh emitter sends a null particle along the outgoing principal null congruence (${\mathcal{L}/\mathcal{E}=a}$), the frequency $\omega_\text{ob}$ seen by the observer simplifies to
\begin{equation}
    \omega_\text{ob}=\dot{t}-a\ \dot{\varphi}=\frac{P_r(r_\text{HO})}{\Delta(r_\text{HO})},
\end{equation}
and the effective temperature of Eq.~(\ref{eq:kappa_chain_HO}) becomes
\begin{equation}\label{eq:kappa_PN_HO}
    \kappa^+=\omega_\text{ob}\frac{\Delta'(r_+)}{2r_+^2}.
\end{equation}
The outgoing effective temperature seen by a horizostationary observer looking in the principal null direction, given by Eq.~(\ref{eq:kappa_PN_HO}), depends only on the black hole's spin-to-mass ratio $a/M$, varying monotonically from ${\kappa^+=1/(4M)}$ when ${a=0}$ to ${\kappa^+=\sqrt{3}/M}$ when ${a=M}$.

Because the effective temperature given by Eq.~(\ref{eq:kappa_PN_HO}) does not change with the observer's proper time, the adiabatic control function $\epsilon^+$ from Eq.~(\ref{eq:epsilon}) is identically zero, so one may be assured that in the geometric optics (high frequency) limit, an inertial observer orbiting at a radius $r_\text{HO}$ will see a Planckian blackbody spectrum of Hawking radiation originating from the direction of the black hole's past horizon.

The horizostationary observer may also look in a variety of other directions along the equatorial plane, by changing the photon angular momentum $\mathcal{L}$ in Eq.~(\ref{eq:omega_em_HO}), to yield a straightforward change in the effective temperature (in accordance with assertion $\#$3 in Sec.~\ref{sec:int}). But regardless of which direction they look along the past horizon, they will always see some non-vacuum state caused by the exponentially redshifting Hawking modes originating from that horizon.

As a final comment, it should be noted that while the horizostationary observer has zero 4-acceleration, they will still experience a twisting force corresponding to non-Fermi-Walker transport. The only non-zero component of their 4-rotation $\mathscr{O}^\mu$ (the angular velocity of their spatial basis vectors with respect to comoving inertial gyroscopes) is the polar component \cite{sem93}
\begin{equation}
    \mathscr{O}^\theta=\frac{\dot{t}^2}{r^6}\left(M\left(1-a\Omega_+\right)\left((3r^2+a^2)\Omega_+-a\right)-r^3\Omega_+\right),
\end{equation}
which is of order unity in the extremal case, falls below 0.1 in black hole mass units when ${a/M\lesssim0.9}$, and vanishes as ${a/M\to0}$. This rotation will in principle induce particle creation via the non-inertial Unruh effect; however, its contribution to the effective temperature calculated in Eq.~(\ref{eq:kappa_PN_HO}) should be negligible for all black hole spins except those near enough to extremality.

\subsection{\label{subsec:ffo}General freely falling observers}

For an arbitrary freely falling observer in the Kerr spacetime, as mentioned in Sec.~\ref{subsec:unr}, they must generally watch a family of Unruh emitters at different angular positions along the black hole's past horizon. To see why this is the case, consider an equatorial infaller observing a single emitter near the past horizon with an angular velocity asymptotically approaching $\Omega_+$ from Eq.~(\ref{eq:Omega_p}). This observer must possess
\begin{equation}
    \frac{d\varphi}{dt}=\Omega_+-\delta\Omega,\qquad\frac{dr}{dt}=-\frac{k^r\delta\Omega}{k^\varphi},
\end{equation}
where ${k^\mu\equiv dx^\mu/d\lambda}$ is the 4-momentum of the null geodesic, and ${\delta\Omega}$ is the observer's differential change in their angular velocity as they fall inwards and ``catch up'' with the azimuthally varying null geodesic. This system of equations has no apparent solution that does not involve position-dependent constants of motion from either the observer or the null ray. Therefore, no infalling observer can both follow a timelike geodesic path and keep up with a single emitter's null ray.

\begin{figure}[t]
\centering
\includegraphics[width=0.75\columnwidth]{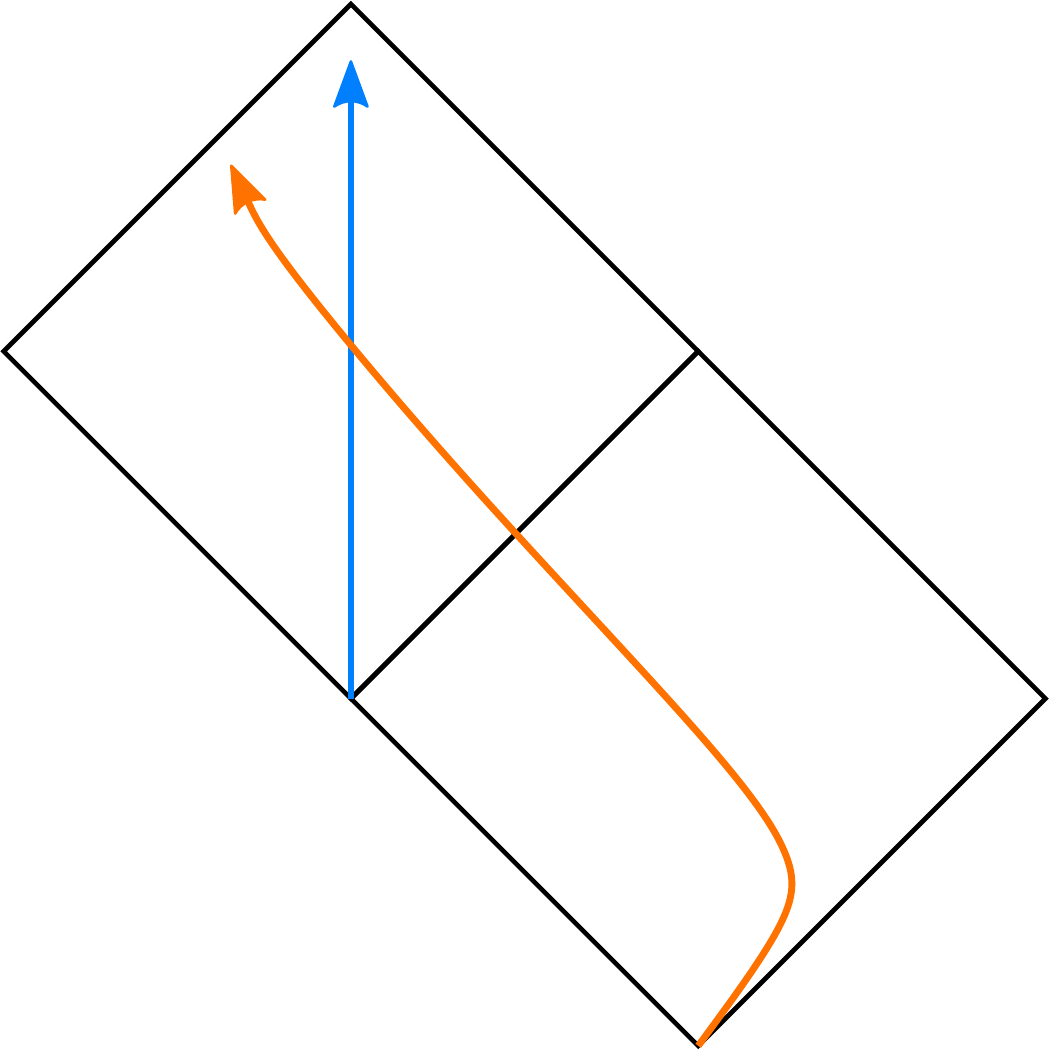}
\caption{Worldlines of the two observers considered in Sec.~\ref{subsec:ffo}: a freely falling equatorial ZAMO (orange path) beginning at rest at infinity and ending at the left portion of the inner horizon, and an interior Carter observer (blue path) beginning at the intersection of the outgoing and ingoing portions of the $r_+$ surface and ending at the intersection of the outgoing and ingoing portions of the $r_-$ surface.\label{fig:penrose_observers}}
\end{figure}

Thus, in calculating the effective temperature seen by an observer watching a family of Unruh emitters along the past horizon, one must impose an additional constraint so that those emitters all lie along the same eikonal wavefront. Namely, the null affine distance, scaled by the frequency measured in the emitter's frame, Eq.~(\ref{eq:lambda_em_ob}), must be held constant. Assuming in what follows that the observer and the emitter remain fixed at the same angular position $\theta$ throughout the course of their trajectories, variations with respect to the observer's proper time from Eq.~(\ref{eq:kappa_lambda_em}) can come only from the observer's and emitter's radial coordinates. Therefore,
\begin{equation}\label{eq:kappa_lambda_em2}
    \kappa=-\dot{r}_\text{ob}\left(\frac{\partial\ln\omega_\text{ob}}{\partial r_\text{ob}}+\frac{\partial\ln\lambda}{\partial r_\text{ob}}\right)-\dot{r}_\text{em}\frac{\omega_\text{ob}}{\omega_\text{em}}\frac{\partial\ln\lambda}{\partial r_\text{em}},
\end{equation}
with the affine distance $\lambda$ given by Eq.~(\ref{eq:lambda}). One additional assumption, as first argued in Ref.~\cite{ham18}, is that the observer should stare in a fixed direction, instead of rotating their frame of reference and inducing non-inertial effects. The direction an observer looks in their field of view can be parameterized by two viewing angles $\chi$ and $\psi$, where ${\chi\in(-\pi,\pi)}$ is the azimuthal angle in their local tetrad frame along the $\gamma_1$-$\gamma_3$ plane (zeroed along the positive $\gamma_1$ axis), and ${\psi\in[0,\pi)}$ is the polar angle from the $\gamma_2$ axis. The viewing angles ${(\chi,\psi)}$, in turn, can be expressed as a function of the 4-momentum of the null particle arriving at the specified point in the observer's field of view, which depends on the observer's position, the emitter's position, and the photon's energy-normalized orbital parameters $\mathcal{L}/\mathcal{E}$ and $\mathcal{K}/\mathcal{E}^2$. In what follows, it is assumed without loss of generality that ${\mathcal{E}=1}$. Then, when the viewing angles ${(\chi,\psi)}$ are kept constant during differentiation, the $\lambda$-dependent terms in Eq.~(\ref{eq:kappa_lambda_em2}) can be expanded with the Leibniz integral rule:
\begin{subequations}\label{eq:dlnlambdadr}
\begin{align}\label{eq:dlnlambdadrob}
    \frac{\partial\ln\lambda}{\partial r_\text{ob}}&=\left(\frac{1}{k^r_\text{ob}}+\frac{\partial\mathcal{L}}{\partial r_\text{ob}}\int_{r_\text{em}}^{r_\text{ob}}\!\!\!dr\ \frac{\partial}{\partial\mathcal{L}}\frac{1}{k^r}\right)/\lambda,\\
    \frac{\partial\ln\lambda}{\partial r_\text{em}}&=-\frac{1}{k^r_\text{em}\lambda}.
\end{align}
\end{subequations}
Eqs.~(\ref{eq:dlnlambdadr}) apply to equatorial geodesics with constant polar coordinate ${\theta=\pi/2}$; the more general case will involve derivatives of both $\mathcal{L}$ and $\mathcal{K}$ applied to the affine distance integrands of Eq.~(\ref{eq:lambda}).

If the photon's angular momentum is large enough that its trajectory contains a turning point, the integration over the affine distance must be split in two, and as a consequence, the derivatives with respect to the constants of motion $\mathcal{L}$ and $\mathcal{K}$ in Eq.~(\ref{eq:dlnlambdadrob}) cannot be brought inside the integral without also introducing a divergent boundary term. In these cases, the derivatives are evaluated numerically with the aid of Richardson extrapolation.

In what follows, two different classes of observers will be considered, as depicted in Fig.~\ref{fig:penrose_observers}. The first is an equatorial observer in free-fall, with zero angular momentum (ZAMO), beginning from rest at infinity. Such an observer has equations of motion given by Eqs.~(\ref{eq:EOM}) with constant of motion ${E=1}$, ${L=0}$, and ${K=a^2}$. The second observer, who can exist only in the interior portion of the black hole, is defined to be at rest in the interior Carter tetrad frame adapted from Eqs.~(\ref{eq:carter}). Such an observer has constants of motion ${E=0}$, ${L=0}$, and ${K=a^2\cos^2\!\theta}$ \cite{mcm21}.

\subsubsection{\label{subsec:ZAMO}Freely falling equatorial ZAMO}

A null particle seen in the locally orthonormal tetrad frame of an observer falling freely with zero angular momentum in the equatorial plane will have the following 4-momentum tetrad components:
\begin{subequations}\label{eq:kobZAMO}
\begin{align}
    k^0_\text{ob}&=\frac{1}{r^2\Delta}\Big(R^2(R^2-a\mathcal{L})+a(\mathcal{L}-a)\Delta+\text{sgn}(k^r)\nonumber\\
    &\times\sqrt{R^2(R^2-\Delta)\left((R^2-a\mathcal{L})^2-(\mathcal{L}-a)^2\Delta\right)}\Big)\label{eq:k0obZAMO}\\
    k^1_\text{ob}&=\frac{R^2k^0_\text{obs}-(R^2-a\mathcal{L})}{\sqrt{R^4-r^2\Delta}}\\
    k^2_\text{ob}&=0\displaybreak[0]\\
    k^3_\text{ob}&=\frac{1}{r\sqrt{R^4-r^2\Delta}}\Big(\left(\mathcal{L}-a\right)\sqrt{R^2(R^2-\Delta)}\nonumber\\
    &-\text{sgn}(k^r)\ a\sqrt{(R^2-a\mathcal{L})^2-(\mathcal{L}-a)^2\Delta}\Big),
\end{align}
\end{subequations}
where, as before, all quantities are normalized to unit photon energy $\mathcal{E}$.

The temporal component $k^0_\text{ob}$ of Eq.~(\ref{eq:k0obZAMO}) is equivalent to the frequency ${\omega_\text{ob}=-k^\mu\dot{x}_\mu}$ seen in the frame of the observer. The spatial components of $k^{\hat{m}}_\text{ob}$ give the angular position of the geodesic in the observer's field of view. This position can be expressed with the viewing angles $\chi$ and $\psi$, defined by
\begin{subequations}\label{eq:chipsi}
\begin{align}
    \tan\chi&\equiv\frac{k^3_\text{ob}}{k^1_\text{ob}},\\
    \tan\psi&\equiv\frac{\sqrt{\left(k^1_\text{ob}\right)^2+\left(k^3_\text{ob}\right)^2}}{k^2_\text{ob}}.
\end{align}
\end{subequations}
The angle $\chi$ gives the observer's azimuthal viewing angle away from the inward radial direction within the equatorial plane, while the polar angle $\psi$ extends to the view out of the plane (here $\psi$ is trivially constant since the observer and emitter are both restricted to the equatorial plane; this condition will be relaxed in the next subsection).

If the observer stares in a fixed direction $\chi$, the null geodesic's angular momentum $\mathcal{L}$ will be found to vary with $r$ as determined from the relation
\begin{equation}\label{eq:coschi}
    \cos\chi=\frac{k^1_\text{ob}}{k^0_\text{ob}}=\frac{R^2-(R^2-a\mathcal{L})(k^0_\text{ob})^{-1}}{\sqrt{R^4-r^2\Delta}}.
\end{equation}

In the Reissner-Nordstr\"om case \cite{mcm23a}, the analog of Eq.~(\ref{eq:coschi}) could be inverted to find an expression for the photon angular momentum $\mathcal{L}$ in terms of the viewing angle $\chi$, so that the effective temperature $\kappa$ could be calculated directly as a function of $\chi$. However, in the present case, no such analytic inversion is possible; instead, the effective temperature will be parameterized by values of $\mathcal{L}$ separately for both ingoing and outgoing photons, and any additional needed quantities like ${d\mathcal{L}/dr_\text{ob}}$ will be found by implicit differentiation of Eq.~(\ref{eq:coschi}).

\begin{figure}[b]
\centering
\includegraphics[width=0.99\columnwidth]{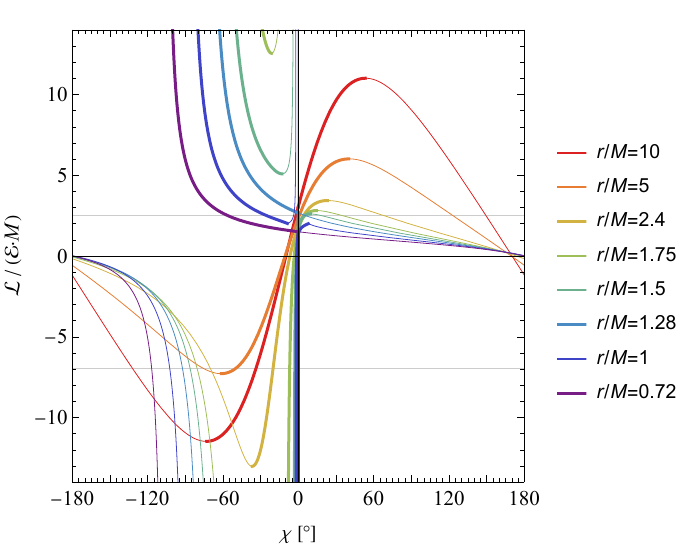}
\caption{Azimuthal viewing angle $\chi$ from Eq.~(\ref{eq:coschi}) for a freely falling equatorial ZAMO as a function of a null geodesic's conserved angular momentum $\mathcal{L}/\mathcal{E}$, for a black hole with spin parameter $a=0.96M$. A selection of different observer radii $r$ are shown, from distant observers (red) to observers crossing the event horizon at $r_+=1.28M$ (blue) to observers crossing the Cauchy horizon at $r_-=0.72M$ (purple). Thick (thin) curves indicate geodesics that are outgoing (ingoing) once they reach the observer. \label{fig:Lofchi_ZAMO}}
\end{figure}

Fig.~\ref{fig:Lofchi_ZAMO} shows the relation between $\mathcal{L}$ and $\chi$ from Eq.~(\ref{eq:coschi}) for observers at various radii when the black hole spin is fixed to ${a/M=0.96}$ (different values of ${a/M}$ yield qualitatively similar plots). For asymptotically distant observers (redder colored curves), the function ${\mathcal{L}(\chi)}$ approaches an exact sinusoid. For a Reissner-Nordstr\"om radial free-faller, this function remains odd for all radii $r$, but for a Kerr ZAMO free-faller, the symmetry is broken by the non-zero spin, so that null geodesics with zero angular momentum are not necessarily aligned with the observer's definition of ${\chi=0^\circ}$.

For reference, the location of the edges of the black hole shadow is indicated in Fig.~\ref{fig:Lofchi_ZAMO} by the intersection of any given curve with the two gray horizontal lines, which lie at the values of $\mathcal{L}$ that solve the equations
\begin{equation}
    \dot{r}=0,\qquad\frac{d\dot{r}}{dr}=0,
\end{equation}
parameterized by the allowed prograde ($-$) and retrograde ($+$) photon orbital radii at the critical values \cite{bar72}
\begin{equation}
    r_c=2M\left(1+\cos\left(\frac{2}{3}\cos^{-1}(\pm a)\right)\right).
\end{equation}
In terms of the photon's orbital parameters $\mathcal{L}$ and $\mathcal{K}$, the edges of the black hole shadow occur at
\begin{subequations}\label{eq:BHsil}
\begin{align}
    \mathcal{L}&=\frac{R^2\Delta'-4r\Delta}{a\Delta'},\\
    \mathcal{K}&=\frac{16r^2\Delta}{(\Delta')^2}.
\end{align}
\end{subequations}

Once the observer is close enough to the black hole to pass within the outermost photon orbit, they begin receiving both outgoing and ingoing photons originating from an emitter just above the event horizon, as shown respectively by the thick and thin portions of each curve in Fig.~\ref{fig:Lofchi_ZAMO}. Then, once the observer falls within the ergosphere bounded by ${r=2M}$, they begin receiving photons with divergent normalized angular momentum $\mathcal{L}/\mathcal{E}$, as shown in Fig.~\ref{fig:Lofchi_ZAMO} by the green curves that dip to negative infinity and reappear in the positive region. In the spherically symmetric case, such divergences happen only below the event horizon and correspond only to a single cusp in $\chi$ instead of a finite swath of $\chi$ values where $\mathcal{L}/\mathcal{E}$ changes sign. Such a rich structure of allowed photon geodesics exists because the Kerr ergosphere extends above the event horizon.

Once the observer reaches the event horizon and proceeds to the inner horizon, the size of the black hole shadow in their field of view remains finite, still governed by the intersections of each colored curve with the two horizontal gray lines in Fig.~\ref{fig:Lofchi_ZAMO}. This black hole shadow marks the position of the past horizon, which sources the Unruh modes contributing to the perception of Hawking radiation.

\begin{figure}[t]
\centering
\includegraphics[width=\columnwidth]{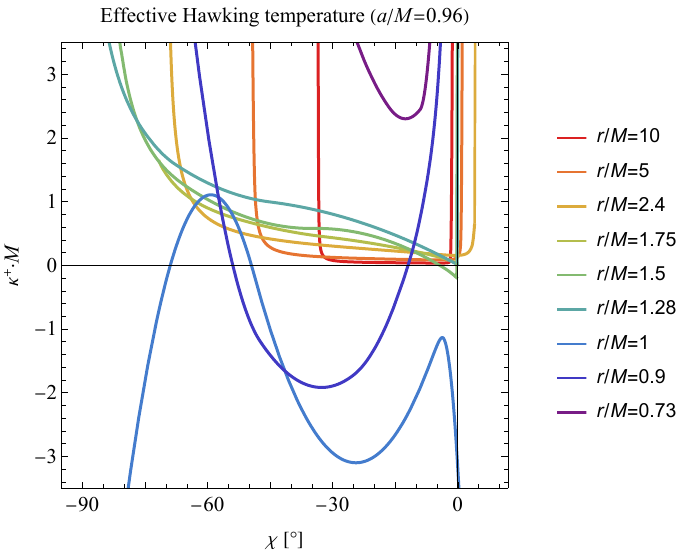}
\caption{Effective Hawking temperature $\kappa^+$ for outgoing Unruh modes as a function of a freely falling equatorial ZAMO's azimuthal viewing angle $\chi$, for a selection of different observer radii $r$. The parameters are identical to that of Fig.~\ref{fig:Lofchi_ZAMO}. \label{fig:kappa_ZAMO}}
\end{figure}

The effective Hawking temperature seen by the freely falling equatorial ZAMO, calculated from Eq.~(\ref{eq:kappa_lambda_em2}), is plotted in Fig.~\ref{fig:kappa_ZAMO} for a selection of observer positions from ${r/M=10}$ down to ${r/M=0.73}$ just above the inner horizon. This temperature depends strongly on the specific choice of observer and exhibits a wide range of behaviors throughout the observer's descent, but a few general trends are worth mentioning.

When the observer is far from the black hole (red curves), the effective temperature is small but non-zero, as expected. As the observer's viewing angle $\chi$ across the equatorial plane changes, so does the effective Hawking temperature, with a minimum value near the center of the black hole shadow and maximum values at the edges. Such behavior is in accordance the limb-brightening assertion $\#$3 in Sec.~\ref{sec:int}, with the only modification that in the Kerr case, the distribution is no longer symmetric about ${\chi=0^{\circ}}$.

As the observer approaches the black hole, the effective Hawking temperature increases in all directions across the black hole shadow, until it becomes negative for certain values of the viewing angle $\chi$. Just as in the case of the on-axis observer of Fig.~\ref{fig:kappa_on-axis}, the effective temperature can be negative even for an observer above the event horizon, as anticipated by assertion $\#$4 in Sec.~\ref{sec:int}.

As the observer approaches the inner horizon, the effective temperature calculated in Fig.~\ref{fig:kappa_ZAMO} diverges to positive infinity (in contrast to the on-axis observer's negative-infinite temperature from Fig.~\ref{fig:kappa_on-axis}). As such, the value of $\kappa^+$ for an observer crossing through the inner horizon at ${r/M=0.72}$ is not shown; instead, the value for an observer just above the inner horizon (${r/M=0.73}$) is displayed, and the effective temperature for any observer closer to the inner horizon will be inversely proportional to the distance above the horizon.

Though not shown explicitly in Fig.~\ref{fig:kappa_ZAMO}, as the effective temperature diverges at ${r\to r_-}$ (when ${r/M<0.73}$ in that plot), the angular distribution across $\chi$ becomes more and more isotropic. Such behavior has been previously noted in both the Schwarzschild \cite{ham18} and Reissner-Nordstr\"om \cite{mcm23a} cases. The key takeaway here and from these prior studies is that the diverging Hawking radiation at the Cauchy horizon is not confined to the single radial point in the observer's field of view where classical radiation diverges via mass inflation, but instead, the diverging semiclassical radiation is distributed uniformly across the entire surface of the black hole's past horizon.

\subsubsection{\label{subsec:CO}Interior Carter observer}

The interior Carter observer, who will also be called the zero-energy observer, is the observer who is at rest in the Carter frame defined by Eqs.~(\ref{eq:carter}) (or, more precisely, with the interior modifications detailed in the text below those equations). This observer moves along the blue path in Fig.~\ref{fig:penrose_observers} and will travel along a constant latitude $\theta$, not necessarily in the equatorial plane as in the previous subsection. In the coordinate frame, the only non-zero component of their 4-velocity is the timelike component ${\dot{r}=-\sqrt{-\Delta}/\rho}$, and in their locally orthonormal tetrad frame, they will see null particles travel with the following 4-momentum components:
\begin{subequations}\label{eq:kobICO}
\begin{align}
    k^0_\text{ob}&=\frac{1}{\rho}\sqrt{\mathcal{K}-\frac{\left(R^2-a\mathcal{L}\right)^2}{\Delta}}\\
    k^1_\text{ob}&=-\frac{R^2-a\mathcal{L}}{\rho\sqrt{-\Delta}}\\
    k^2_\text{ob}&=\frac{\text{sgn}(k^\theta)}{\rho}\sqrt{\mathcal{K}-\frac{\left(\mathcal{L}-a\sin^2\!\theta\right)^2}{\sin^2\!\theta}}\\
    k^3_\text{ob}&=-\frac{\mathcal{L}-a\sin^2\!\theta}{\rho\sin\theta}.
\end{align}
\end{subequations}

As in the previous subsection, the temporal component $k^0_\text{ob}$ of Eqs.~(\ref{eq:kobICO}) is equivalent to the frequency ${\omega_\text{ob}=-k^\mu\dot{x}_\mu}$ seen in the frame of the observer, and the spatial components of $k^{\hat{m}}_\text{ob}$ give the angular position of the geodesic in the observer's field of view, parameterized by the viewing angles $\chi$ and $\psi$ defined by Eqs.~(\ref{eq:chipsi}).

Eqs.~(\ref{eq:chipsi}) can be solved for the photons of Eqs.~(\ref{eq:kobICO}) to yield the following relations between the viewing angles ($\chi$, $\psi$) and the photon's orbital parameters ($\mathcal{L}$, $\mathcal{K}$):
\begin{subequations}\label{eq:chipsiICO}
\begin{align}
    \mathcal{L}&=\frac{a\sin^2\!\theta\sqrt{-\Delta}+R^2\sin\theta\tan\chi}{\sqrt{-\Delta}+a\sin\theta\tan\chi},\\
    \mathcal{K}&=\frac{\rho^4\left(\sec^2\!\chi\csc^2\!\psi-1\right)}{\left(\sqrt{-\Delta}+a\sin\theta\tan\chi\right)^2}.
\end{align}
\end{subequations}

Eqs.~(\ref{eq:chipsiICO}) can be used together with Eq.~(\ref{eq:kappa_lambda_em2}) to calculate the effective Hawking temperature $\kappa$ directly as a function of the observer's azimuthal and polar viewing angles $\chi$ and $\psi$, respectively. Before presenting the results, two modifications from the previous subsection are worth noting. First, the integration of Eq.~(\ref{eq:lambda}) to calculate the affine distance will now include both $r$-dependent and $\theta$-dependent terms, since the observer can now look outside of the equatorial plane. Derivatives with respect to both $\mathcal{L}$ and $\mathcal{K}$ must then be applied to both the $r$-dependent and $\theta$-dependent integrands, in contrast to the simpler case of Eq.~(\ref{eq:dlnlambdadrob}).

Second, while the emitter's radius $r_\text{em}$ can always be fixed at the value $r_+$ (for $\kappa^+$) or $\infty$ (for $\kappa^-$), the emitter's polar angle $\theta_\text{em}$ will change for different values of the photon's orbital parameters $\mathcal{L}$ and $\mathcal{K}$. It must therefore be calculated via the same ray-tracing techniques used throughout this section. If the back-propagated null geodesic originating from an equatorial observer exceeds a Mino time $\tilde{\lambda}$ [defined in Eq.~(\ref{eq:mino})] of
\begin{equation}
    \tilde{\lambda}>K\left(\frac12+\frac{a(a-\mathcal{L})-\mathcal{K}}{\sqrt{(\mathcal{K}+4a\mathcal{L})\mathcal{K}}}\right)\left((\mathcal{K}+4a\mathcal{L})\mathcal{K}\right)^{-1/4},
\end{equation}
where $K$ is the complete elliptic integral of the first kind, then the photon will experience a turning point at
\begin{equation}
    \sin\theta=\frac{\sqrt{\mathcal{K}+4a\mathcal{L}}-\sqrt{\mathcal{K}}}{2a}.
\end{equation}
Since the emitter's polar angle $\theta_\text{em}$ will vary as the observer varies their proper time $\tau_\text{ob}$ while staring in the same fixed direction parametrized by angles ${(\chi,\psi)}$, Eq.~(\ref{eq:kappa_lambda_em2}) will in principle include additional terms with derivatives with respect to $\theta_\text{em}$, even if both the observer's and emitter's polar velocities $\dot{\theta}_\text{ob}$ and $\dot{\theta}_\text{em}$ are individually assumed to be zero (as is the case here). To address this complication, the dependence of $\theta_\text{em}$ on $r_\text{ob}$ and $r_\text{em}$ is explicitly included when evaluating the $r$ derivatives of Eq.~(\ref{eq:kappa_lambda_em2}).

\begin{figure}[t]
\centering
\includegraphics[width=0.99\columnwidth]{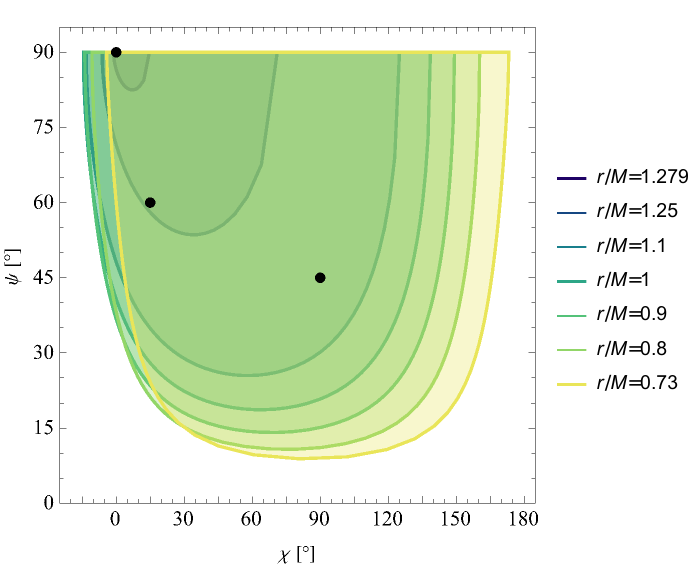}
\caption{Fields of view of the black hole shadow, with edges parametrized by Eqs.~(\ref{eq:BHsil}), for the viewing angles $\chi$ and $\psi$ defined by Eqs.~(\ref{eq:chipsi}). The view is from the perspective of a freely falling equatorial interior zero-energy observer at different radii $r$ within a Kerr black hole with spin parameter ${a=0.96M}$. The shadow initially appears as an infinitesimally small point at $(\chi,\psi)=(0^{\circ},90^{\circ})$ when the observer is at the event horizon (dark blue), then grows along both angular directions as the observer approaches the inner horizon (yellow). The three black points correspond to the three curves shown in Fig.~\ref{fig:kappa_ICO}. \label{fig:chipsi_ICO}}
\end{figure}

\begin{figure}[t]
\centering
\includegraphics[width=0.92\columnwidth]{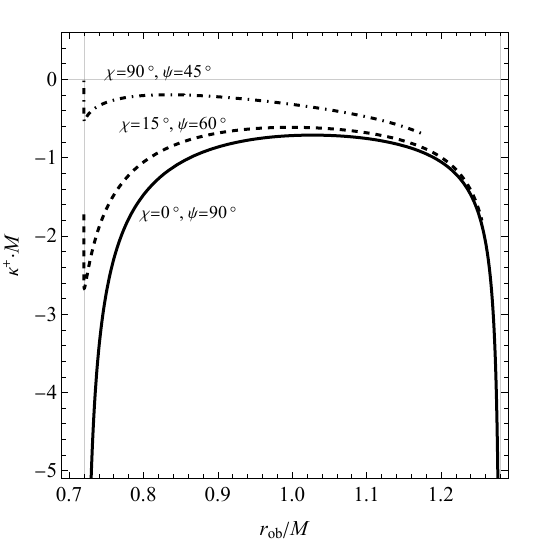}
\caption{Effective Hawking temperature $\kappa^+$ for outgoing Unruh modes seen by a freely falling zero-energy equatorial observer in the interior of a Kerr black hole with spin parameter ${a=0.96M}$. The three curves show the view in the three specific directions labeled in the plot and marked by the black points in Fig.~\ref{fig:kappa_ICO}. \label{fig:kappa_ICO}}
\end{figure}

A freely falling observer in the interior of a Kerr black hole with zero energy will see the black hole shadow grow over time, as shown in Fig.~\ref{fig:chipsi_ICO}. As the interior Carter observer begins at ${r_\text{ob}=r_+}$ at the bifurcation point of the past horizon and the event horizon, they initially see the black hole shadow emerge from a single point in their field of view along the principal null direction at ${\chi=0^\circ}$, ${\psi=90^\circ}$ (the upper-left point in Fig.~\ref{fig:chipsi_ICO}). Then, the black hole shadow appears to grow in their field of view until taking on the yellow shape in Fig.~\ref{fig:chipsi_ICO} when the observer reaches the inner horizon. These are the regions that appear as the source of outgoing Hawking modes in the observer's field of view.

\begin{figure}[t]
\centering
\includegraphics[width=\columnwidth]{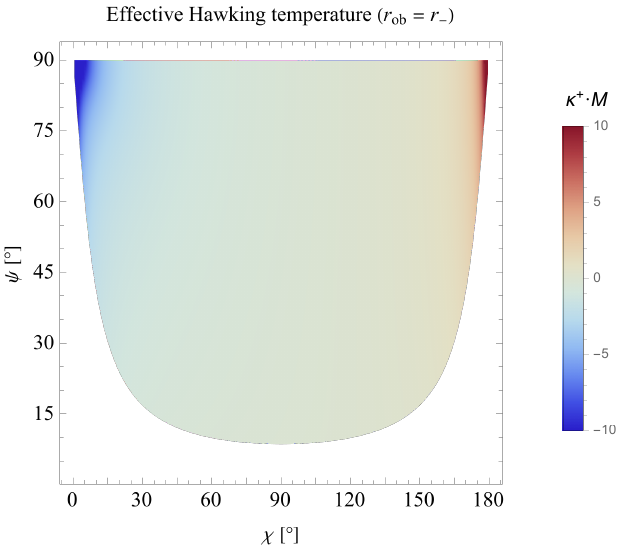}
\caption{Effective Hawking temperature $\kappa^+$ seen by an equatorial interior zero-energy observer just above the inner horizon of a Kerr black hole (with spin parameter ${a=0.96M}$). The effective temperature is mostly (but not completely) isotropic and diverges to $\pm\infty$ along the principal null directions. \label{fig:kappa_r-_ICO}}
\end{figure}

The calculation of the effective Hawking temperature $\kappa^+$ via Eq.~(\ref{eq:kappa_lambda_em2}) for an interior Carter observer, who follows the blue path in Fig.~\ref{fig:penrose_observers}, is presented in Figs.~\ref{fig:kappa_ICO} and \ref{fig:kappa_r-_ICO}. In Fig.~\ref{fig:kappa_ICO}, three specific viewing directions are chosen to track how $\kappa^+$ changes as the observer travels from the event horizon to the inner horizon. These three directions are denoted by the black points in Fig.~\ref{fig:chipsi_ICO}: at the approximate center of the shadow at $(\chi,\psi)=(90^\circ,45^\circ)$, closer to the edge at $(\chi,\psi)=(15^\circ,60^\circ)$, and at the point of emergence at $(\chi,\psi)=(0^\circ,90^\circ)$.

When staring along the three directions shown in Fig.~\ref{fig:kappa_ICO}, the observer sees a wide range of effective temperatures. The temperature appears to converge to a negative, infinite value as ${r_\text{ob}\to r_+}$, since at this point, the observer is coincident with the past horizon singularity imposed by the Unruh vacuum state. However, most directions the observer might look in the sky (in fact, all but a set of zero measure) do not actually reach this pathological divergence, since the black hole shadow falls out of their range above a certain radius.

As the interior Carter observer approaches the inner horizon, the effective temperature does not diverge in every direction, as it did for the equatorial ZAMO in the previous Sec.~\ref{subsec:ZAMO} and for the Reissner-Nordstr\"om radial infallers of Ref.~\cite{mcm23a}. Instead, $\kappa^+$ approaches a finite value in every direction along the black hole shadow except along the principal null directions at ${\chi=0^\circ}$ and ${\chi=180^\circ}$, where $\kappa^+$ does diverge to $-\infty$ and $+\infty$, respectively. One of these divergences (${\chi=0^\circ,\psi=90^\circ}$) is shown in Fig.~\ref{fig:kappa_ICO}.

The full view of the effective Hawking temperature seen just above the inner horizon is shown in Fig.~\ref{fig:kappa_r-_ICO}. The effective temperature becomes approximately isotropic and negligibly small across most of the surface of the past horizon, but it diverges to ${\pm\infty}$ and exceeds the saturation limit of the color scale along the ingoing and outgoing principal null directions at the top left and top right corners of the figure.

In conclusion, the effective temperature of Hawking radiation can be calculated in the geometric optics framework for any class of inertial observers within the Kerr spacetime, with widely varying outcomes depending on the particular choice of orbital parameters and spacetime positions. In this Sec.~\ref{sec:eff}, we have examined four such classes of observers: freely falling observers along the axis of rotation, observers in a horizostationary orbit, equatorial infallers with zero angular momentum, and equatorial infallers with zero energy. In the former two cases, an effective temperature could be calculated purely as the rate of redshift between the observer and a single freely falling emitter in the Unruh state, while in the latter two cases, an additional constraint that the emitted affine distance be kept constant was required so that a family of Unruh-state emitters could be matched to the same eikonal wavefront as the observer pans across their field of view.

For all classes of observers examined here that reach the black hole's Cauchy horizon, at least one point in their field of view contains a diverging effective Hawking temperature. In accordance with prior studies of both Hawking radiation \cite{ham18,mcm23a} and the renormalized stress-energy tensor \cite{zil22b}, this semiclassically divergent behavior appears to be generic. Though here we have not proved that Hawking radiation temperatures will diverge for every inner-horizon observer within the Kerr spacetime, the fact that a divergence appears for even a single inertial observer is enough to demonstrate that Kerr black holes are semiclassically singular and likely unstable at the inner horizon.

\section{\label{sec:bog}Bogoliubov spectrum}

The results of the previous section indicate that a Kerr infaller observing at high frequencies will see an approximately thermal spectrum of Hawking radiation throughout their inward journey, until the temperature of that radiation diverges while looking in at least one direction as they approach the inner horizon. While these results are robust in specific instances, they are confined to regimes in which the observer satisfies the adiabatic condition of Eq.~(\ref{eq:epsilon}), and further, they are not guaranteed to provide any information about the behavior of Hawking radiation below the high-frequency geometric optics limit.

In order to address these concerns, consider the set of limiting cases in which the simplifying assumptions of the effective temperature formalism can be circumvented and the full spectrum of Hawking radiation can be calculated directly from Eq.~(\ref{eq:bog}).

\subsection{\label{subsec:bogder}Derivation}
The focus of the present section will be the cases in which the scattering potential of the radial Klein-Gordon\footnote{The calculations of the Hawking spectra, and in particular the formulae of Eqs.~(\ref{eq:N}), are valid for any bosonic field with integer spin, with the only change coming from the numerically-obtained values of the scattering coefficients for a given field's wave equation; see Appendix~\ref{app:mst} for more details. In the derivation that follows, focus will be placed on the scalar (spin-0) case.} wave equation, given by Eq.~(\ref{eq:V}), asymptotically reduces to a constant in the tortoise coordinate $r^*$. Such cases occur when the observer is asymptotically far away and when the observer is crossing one of the black hole's horizons:
\begin{equation}\label{eq:V_asymp}
    V_{\omega\ell m}\to
    \begin{cases}
        \omega,&r\to\infty\\
        \omega_\pm,&r\to r_\pm
    \end{cases},
\end{equation}
where
\begin{equation}\label{eq:omega_pm}
    \omega_\pm\equiv\omega-m\Omega_\pm=\omega-m\frac{a}{R_\pm^2}.
\end{equation}
In these limits, the wave equation possesses asymptotic eigenmode solutions of the form ${\text{exp}(\pm iV_{\omega\ell m}r^*)}$, and the problem of mode propagation between these limits reduces to a 1D scattering problem in $r^*$.

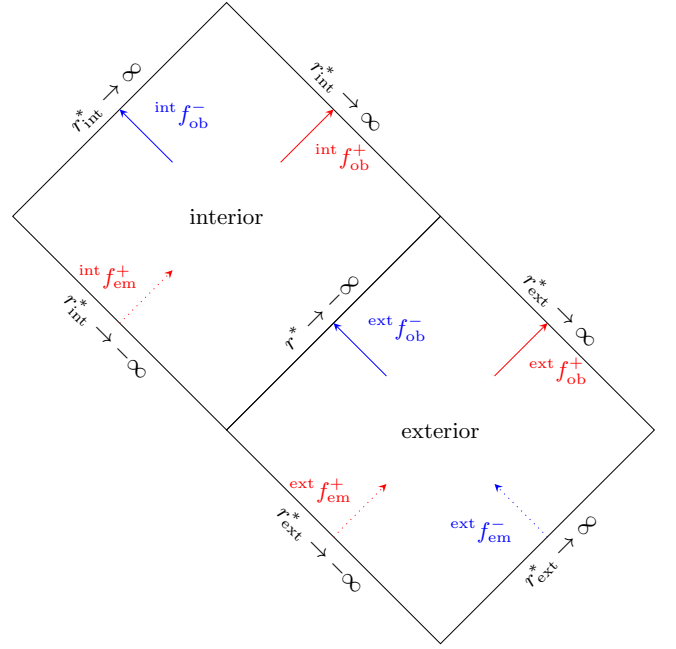
\begin{figure}[t]
\resizebox{\columnwidth}{!}{%
\begin{tikzpicture}
\def\x{1.5cm} 
\node (ext)   at (\x,-\x) {exterior};
\node (int)   at (-\x,\x) {interior};
\path  
  (int) +(90:2*\x)  coordinate  (inttop)
        +(-90:2*\x) coordinate  (intbot)
        +(0:2*\x)   coordinate  (intright)
        +(180:2*\x) coordinate  (intleft);
\draw (intleft) -- 
          node[midway, above left, rotate=45, anchor=south] {$r^*_\text{int}\to\infty$}
      (inttop) --
          node[midway, above right, rotate=-45, anchor=south] {$r^*_\text{int}\to\infty$}
      (intright) -- (intbot) --
          node[midway, below left, rotate=-45, anchor=north] {$r^*_\text{int}\to-\infty$}
      (intleft) -- cycle;
\path 
   (ext) +(90:2*\x)  coordinate
            (exttop)
         +(-90:2*\x) coordinate
            (extbot)
         +(180:2*\x) coordinate
            (extleft)
         +(0:2*\x)   coordinate
            (extright);
\draw  (extleft) -- 
          node[midway, above left, rotate=45, anchor=south] {$r^*\to-\infty$}
    (exttop) --
          node[midway, above right, rotate=-45, anchor=south] {$r^*_\text{ext}\to\infty$}
    (extright) -- 
          node[midway, below right, rotate=45, anchor=north] {$r^*_\text{ext}\to\infty$}
    (extbot) --
          node[midway, below left, rotate=-45, anchor=north] {$r^*_\text{ext}\to-\infty$}
    (extleft) -- cycle;
\draw [dotted,red,-stealth](0,-2*\x) -- 
          node[midway, above left] {${}^\text{ext}f^+_\text{em}$}
    +(45:0.7*\x);
\draw [dotted,red,-stealth](-2*\x,0) --
          node[midway, above left] {${}^\text{int}f^+_\text{em}$}
    +(45:0.7*\x);
\draw [dotted,blue,-stealth](2*\x,-2*\x) --
          node[midway, below left] {${}^\text{ext}f^-_\text{em}$}
    +(135:0.7*\x);
\draw [red,stealth-](2*\x,0) -- 
          node[midway, below right] {${}^\text{ext}f^+_\text{ob}$}
    +(-135:0.7*\x);
\draw [red,stealth-](0,2*\x) --
          node[midway, below right] {${}^\text{int}f^+_\text{ob}$}
    +(-135:0.7*\x);
\draw [blue,stealth-](-2*\x,2*\x) --
          node[midway, above right] {${}^\text{int}f^-_\text{ob}$}
    +(-45:0.7*\x);
\draw [blue,stealth-](0,0) --
          node[midway, above right] {${}^\text{ext}f^-_\text{ob}$}
    +(-45:0.7*\x);
\end{tikzpicture}
}%
\caption{Penrose diagram showing the various boundaries for a Kerr black hole on which modes are given non-zero initial data. Superscripts $+$ ($-$) everywhere indicate whether modes traveling across a surface are outgoing (ingoing). The initial data for the emitter's (observer's) modes at the locations of the dotted (solid) lines can then be propagated (back-propagated) numerically using the wave equation to define the modes throughout the entire spacetime.\label{fig:penrose}}
\end{figure}

One can define the following future boundary conditions (i.e., in the limit as the timelike coordinates $t_\text{ext}$ or $r^*_\text{int}$ are taken to positive infinity in their respective domains) for each of four complete sets of radial mode solutions to the wave Eq.~(\ref{eq:waveeq_r}), corresponding to observers locally defining a positive frequency $\omega$ and azimuthal quantum number $m$:
\begin{subequations}\label{eq:f_ob_future}
\begin{align}
    {}^{\text{ext}}f_\text{ob}^+&\to
    \begin{cases}
        \text{e}^{im\varphi-i\omega u},&r^*_\text{ext}\to\infty\\
        0,&r^*_\text{ext}\to-\infty
    \end{cases},\label{eq:f_ob_future_pext}\\
    {}^{\text{ext}}f_\text{ob}^-&\to
    \begin{cases}
        \text{e}^{im\varphi_+-i\omega v},&r^*_\text{ext}\to-\infty\\
        0,&r^*_\text{ext}\to\infty
    \end{cases},\label{eq:f_ob_future_mext}\\
    {}^{\text{int}}f_\text{ob}^+&\to
    \begin{cases}
        \text{e}^{-im\varphi_-+i\omega u},&t_\text{int}\to\infty\\
        0,&t_\text{int}\to-\infty\ \cup\ r^*_\text{ext}\to\infty
    \end{cases},\label{eq:f_ob_future_pint}\\
    {}^{\text{int}}f_\text{ob}^-&\to
    \begin{cases}
        0,&t_\text{int}\to\infty\ \cup\ r^*_\text{ext}\to\infty\\
        \text{e}^{im\varphi_--i\omega v},&t_\text{int}\to-\infty\label{eq:f_ob_future_mint}
    \end{cases},
\end{align}
\end{subequations}
where $\varphi_\pm$, defined in Eq.~(\ref{eq:phi_pm}), is the azimuthal coordinate that is regular for an observer crossing the horizon at ${r=r_\pm}$. Note that these modes differ slightly from the Eddington-Finkelstein modes used in Ref.~\cite{zil22a} in the use of $\omega$ rather than $\omega_\pm$ at the outer/inner horizons, since the modes here are constructed explicitly to match the positive-frequency experience of a free-falling observer rather than to provide pure eigenmode solutions to the wave equation (more details below). These initialized modes are shown by the solid arrows in Fig.~\ref{fig:penrose}. The notation for labeling these modes is the same as in Ref.~\cite{mcm23a}; modes in the exterior (interior) portion of the spacetime are labeled ${{}^\text{ext}f}$ (${{}^\text{int}f}$), and modes with canonically affine boundary conditions along a future null boundary transverse to outgoing (ingoing) null rays are labeled $f^+$ ($f^-$).

The subscript ``ob'' in the modes of Eqs.~(\ref{eq:f_ob_future}) is used to indicate that each mode corresponds to the waves that would be seen in the frame of an inertial observer positioned asymptotically close to its respective null boundary. To see why this is the case, consider the following analysis in analog to that performed in Sec.~\ref{subsec:unr} for the modes of the emitter.

If an infalling observer is placed asymptotically far from the black hole at rest, with ${E=1}$, the outgoing modes encoded by ${{}^\text{ext}f_\text{ob}^+}$ will track the same eikonal wavefront as the outgoing null congruence derived from their own proper time, since
\begin{equation}
    \lim_{r\to\infty}\frac{du}{d\tau}=\lim_{r\to\infty}\left(\dot{t}-\frac{R^2}{\Delta}\dot{r}\right)=1.
\end{equation}
Similarly, if an infalling observer is placed asymptotically close to the event horizon, they will see the ingoing waves of ${{}^\text{ext}f_\text{ob}^-}$ tick at a rate proportional to their own proper time:
\begin{equation}\label{eq:vdot_r+}
    \lim_{r\to r_+}\frac{dv}{d\tau}=\frac{aP_\theta}{\rho^2}+\frac{R^2(K+r^2)}{2\rho^2P_r},
\end{equation}
with the Hamilton-Jacobi parameters ${P_r(r)>0}$ and ${P_\theta(\theta)}$ defined in Eq.~(\ref{eq:Prth}). Note that while the expression on the right-hand side of Eq.~(\ref{eq:vdot_r+}) does not generally reduce to unity (although it does simplify to ${1/2}$ for the on-axis observer considered in Sec.~\ref{subsubsec:on-axis}), the expression is nonetheless frozen at a constant value as $r^*$ is varied, in contrast to the divergent behavior for \emph{outgoing} waves seen by an ingoing emitter at the event horizon from Eq.~(\ref{eq:udot_r+}).

For the remaining two interior modes ${{}^\text{int}f_\text{ob}^\pm}$, if an infalling (${\dot{r}<0}$) observer is placed at the inner horizon and is ingoing (${P_r>0}$), their proper time will be proportional to the ingoing modes ${{}^\text{int}f_\text{ob}^-}$, while if the observer is outgoing (${P_r<0}$), their proper time will be proportional to the outgoing modes ${{}^\text{int}f_\text{ob}^+}$. The constant of proportionality is the same as the right-hand side of Eq.~(\ref{eq:vdot_r+}), with $r$ now at its inner horizon value. Thus, the modes defined by Eqs.~(\ref{eq:f_ob_future}) each correspond to the eikonal waves seen by an inertial, infalling observer passing through their respective hypersurface boundaries.

Since each set of boundaries considered in Eqs.~(\ref{eq:f_ob_future}) for each of the four sets of modes forms a complete null Cauchy hypersurface terminating at spacelike infinity, the radial wave Eq.~(\ref{eq:waveeq_r}) can be used to back-propagate each mode throughout the rest of the spacetime. Of particular importance is the behavior of these observer modes at the past null boundaries where the initial data for the emitter's Unruh modes are defined, since if both $f_\text{ob}$ and $f_\text{em}$ are known along the same Cauchy hypersurface, Eqs.~(\ref{eq:bog}) and (\ref{eq:inner_product}) can be used to compute the scalar product between the observer's and emitter's modes and therefore the spectrum of Hawking radiation.

Equivalently, one may consider propagating the emitter's modes forward and evaluating the mode scalar product along the future null boundary where the initial data for the observer's modes are defined, instead of propagating the observer's modes backward to the past null boundary. However, this task is more difficult since the Kruskal coordinate $U$ used to define the emitter's Unruh modes contains non-trivial coupling between $t$ and $r$, so that the wave equation is not separable in these coordinates when initialized with the Unruh modes.

Fortunately, as mentioned above, the problem of finding the observer modes at the past null boundaries is a straightforward 1D scattering problem in $r^*$. Define reflection coefficients $\mathcal{R}^\pm_{\text{int,ext}}$ and transmission coefficients $\mathcal{T}^\pm_{\text{int,ext}}$ for the interior and exterior portions of the spacetime (with the same notation as in Ref.~\cite{mcm23a}), depicted visually by the scattering paths in the Penrose diagrams to the right of each of the expressions below. The boundary conditions to be solved for the radial modes of Eq.~(\ref{eq:modesep_f}), evaluated at the mode labeled with $\omega$, $\ell$, and $m$, are provided in Eqs.~(\ref{eq:f_obext+})\textendash(\ref{eq:f_obint-}) below. For the modes of Eq.~(\ref{eq:f_ob_future_pext}) encoded by an observer asymptotically far away from the black hole, one has
\begin{widetext}
\begin{equation}\label{eq:f_obext+}
    {}^\text{ext}\psi^+_{\text{ob}}\to
    \begin{cases}
        \text{e}^{i\omega r^*}+\mathcal{R}^+_{\text{ext},\omega}\text{e}^{-i\omega r^*},&r^*_\text{ext}\to\infty,\phantom{-}\qquad\Rpext\\
        \mathcal{T}^+_{\text{ext},\omega}\text{e}^{i\omega_+r^*},&r^*_\text{ext}\to-\infty,\qquad\Tpext
    \end{cases},
\end{equation}
for the modes of Eq.~(\ref{eq:f_ob_future_mext}) encoded by an ingoing observer at the event horizon, one has
\begin{equation}\label{eq:f_obext-}
    {}^\text{ext}\psi^-_\text{ob}\to
    \begin{cases}
        \mathcal{T}^-_{\text{ext},\omega}\text{e}^{-i(\omega+m\Omega_+)r^*},&r^*_\text{ext}\to\infty,\phantom{-}\qquad\Tmext\\
        \text{e}^{-i\omega r^*}+\mathcal{R}^-_{\text{ext},\omega}\text{e}^{i\omega r^*},&r^*_\text{ext}\to-\infty,\qquad\Rmext
    \end{cases},
\end{equation}
for the modes of Eq.~(\ref{eq:f_ob_future_pint}) encoded by an outgoing observer at the inner horizon, one has
\begin{equation}\label{eq:f_obint+}
    {}^\text{int}\psi^+_\text{ob}\to
    \begin{cases}
        \text{e}^{-i\omega r^*},&r^*_\text{int}\to\infty,\\
        \mathcal{T}^+_{\text{int},\omega}\text{e}^{-i(\omega_++m\Omega_-)r^*}+\mathcal{R}^+_{\text{int},\omega}\text{e}^{i(\omega_++m\Omega_-)r^*},&r^*_\text{int}\to-\infty,\qquad\Tpint+\Rpint\\
        \mathcal{R}^+_{\text{int},\omega}\left(\text{e}^{i(\omega_++m\Omega_-)r^*}+\mathcal{R}^-_{\text{ext},\omega}\text{e}^{-i(\omega_++m\Omega_-)r^*}\right),&r^*_\text{ext}\to-\infty,\qquad\Rpint+\RmextRpint\\
        \mathcal{R}^+_{\text{int},\omega}\mathcal{T}^-_{\text{ext},\omega}\text{e}^{i(\omega+m\Omega_-)r^*},&r^*_\text{ext}\to\infty,\phantom{-}\qquad\TmextRpint
    \end{cases},
\end{equation}
and for the modes of Eq.~(\ref{eq:f_ob_future_mint}) encoded by an ingoing observer at the inner horizon, one has
\begin{equation}\label{eq:f_obint-}
    {}^\text{int}\psi^-_\text{ob}\to
    \begin{cases}
        \text{e}^{-i\omega r^*},&r^*_\text{int}\to\infty,\\
        \mathcal{T}^-_{\text{int},\omega}\text{e}^{-i(\omega_++m\Omega_-)r^*}+\mathcal{R}^-_{\text{int},\omega}\text{e}^{i(\omega_++m\Omega_-)r^*},&r^*_\text{int}\to-\infty,\qquad\Tmint+\Rmint\\
        \mathcal{T}^-_{\text{int},\omega}\left(\text{e}^{-i(\omega_++m\Omega_-)r^*}+\mathcal{R}^-_{\text{ext},\omega}\text{e}^{i(\omega_++m\Omega_-)r^*}\right),&r^*_\text{ext}\to-\infty,\qquad\Tmint+\RmextTmint\\
        \mathcal{T}^-_{\text{int},\omega}\mathcal{T}^-_{\text{ext},\omega}\text{e}^{-i(\omega+m\Omega_-)r^*},&r^*_\text{ext}\to\infty,\phantom{-}\qquad\TmextTmint
    \end{cases}.
\end{equation}

\end{widetext}

The scattering coefficients in the above expressions can be calculated through numerical means; here we use the Teukolsky 0.3.0 package of the Black Hole Perturbation Toolkit \cite{bhp}. Since this Mathematica package is only designed to compute exterior scattering coefficients, we have made adaptations to the code to extend computations to the spacetime region between the inner and outer horizons; details on these modifications can be found in Appendix~\ref{app:mst}.

The scalar product of Eq.~(\ref{eq:inner_product}) can then be evaluated along the past null Cauchy hypersurface where the Unruh state is initialized. The end goal is the computation of Eq.~(\ref{eq:bog}), the vacuum expectation value of the particle number operator for an observer either at infinity, the event horizon, or the ingoing or outgoing portions of the inner horizon. These spectral number distributions will be labeled ${\langle N^+_\text{ext}\rangle_{\omega\ell m}}$, ${\langle N^-_\text{ext}\rangle_{\omega\ell m}}$, ${\langle N^+_\text{int}\rangle_{\omega\ell m}}$, and ${\langle N^-_\text{int}\rangle_{\omega\ell m}}$ for the respective modes of Eqs.~(\ref{eq:f_obext+})\textendash(\ref{eq:f_obint-}).

The analysis proceeds almost identically to that of the spherical case in Ref.~\cite{mcm23a}, with one small but crucial difference: the Kerr scattering potential of Eq.~(\ref{eq:V}) asymptotically approaches a different constant value at infinity compared to the values at the event horizon and the inner horizon; see Eq.~(\ref{eq:V_asymp}). Thus, observer modes that are initialized as 
\begin{equation}
    \phi_\text{ob}\sim\frac{\text{e}^{i\omega r^*}}{R\sqrt{4\pi\omega}}
\end{equation}
at future null infinity will be back-scattered into the form
\begin{equation}
    \phi_\text{ob}\sim\frac{\text{e}^{i\omega_+r^*}}{R_+\sqrt{4\pi\omega}}
\end{equation}
along the past horizon, and so forth.

The details for the calculation of the resulting number operator vacuum expectation values from Eq.~(\ref{eq:bog}) are given in Appendix~\ref{app:bog}. The result, up to a normalization factor, is
\begin{widetext}
\begin{subequations}\label{eq:N}
\begin{align}
    \langle N^+_\text{ext}\rangle_{\omega\ell m}&=\left(\frac{\omega_+}{\omega}\right)\frac{\left|\mathcal{T}^+_{\text{ext},\omega}\right|^2}{\text{e}^{\frac{2\pi}{\varkappa_+}\omega_+}-1},\displaybreak[0]\label{eq:N+ext}\\
    \langle N^-_\text{ext}\rangle_{\omega\ell m}&=\frac{\left|\mathcal{R}^-_{\text{ext},\omega}\right|^2}{\text{e}^{\frac{2\pi}{\varkappa_+}\omega}-1},\displaybreak[0]\label{eq:N-ext}\\
    \langle N^-_\text{int}\rangle_{\omega\ell m}&=\left(\frac{\omega_++m\Omega_-}{\omega}\right)\frac{\left|\mathcal{T}^-_{\text{int},\omega}\mathcal{R}^-_{\text{ext},\omega}-\mathcal{R}^-_{\text{int},\omega}\text{e}^{\frac{\pi}{\varkappa_+}(\omega_++m\Omega_-)}\right|^2}{\text{e}^{\frac{2\pi}{\varkappa_+}(\omega_++m\Omega_-)}-1},\label{eq:N-int}\\
    \langle N^+_\text{int}\rangle_{\omega\ell(-m)}&=\left(\frac{\omega_++m\Omega_-}{\omega}\right)\frac{\left|\mathcal{T}^+_{\text{int},\omega}-\mathcal{R}^+_{\text{int},\omega}\mathcal{R}^-_{\text{ext},\omega}\text{e}^{\frac{\pi}{\varkappa_+}(\omega_++m\Omega_-)}\right|^2}{\text{e}^{\frac{2\pi}{\varkappa_+}(\omega_++m\Omega_-)}-1}+\left(\frac{\omega+m\Omega_-}{\omega}\right)\left|\mathcal{R}^+_{\text{int},\omega}\mathcal{T}^-_{\text{ext},\omega}\right|^2,\label{eq:N+int}
\end{align}
\end{subequations}
\end{widetext}
with the surface gravity $\varkappa_+$ of Eq.~(\ref{eq:surf}), the horizon-limit frequency $\omega_+$ of Eq.~(\ref{eq:omega_pm}), and the transmission and reflection coefficients $\mathcal{T}^\pm_{\text{int,ext}}$ and $\mathcal{R}^\pm_{\text{int,ext}}$ of Eqs.~(\ref{eq:f_obext+})\textendash(\ref{eq:f_obint-}). Note that in the limit ${a\to0}$, the expressions in Eqs.~(\ref{eq:N}) reduce to the same Schwarzschild expressions obtained in Ref.~\cite{mcm23a}.

The physical interpretation of the vacuum expectation values from Eqs.~(\ref{eq:N}) is that they measure the spectral emission rate; i.e., the mean number of quanta in the mode with frequency $\omega$ and angular mode numbers $\ell$ and $m$, seen by a freely falling observer at each respective location, per the observer's proper time $\tau_\text{ob}$.

One may at first sight worry that for exterior scattering, when ${\omega<m\Omega_+}$ (and similarly, ${\omega<m(\Omega_+-\Omega_-)}$ for the interior), the frequency prefactors in the above expressions for ${\langle N_\text{ext}^+\rangle_{\omega\ell m}}$ and ${\langle N_\text{int}^\pm\rangle_{\omega\ell m}}$ become negative. This is connected to the well-known phenomenon of superradiance, in which the transmission probability in a rotating system becomes negative and the absorption probability exceeds unity, so that scattered waves gain amplitude upon reflection and extract energy from the black hole \cite{sta73}. However, the aforementioned negative terms are exactly canceled by the Planckian terms in the denominator of each expression, which also become negative in the same superradiant regimes. Therefore, the expected number of particles seen by the observer will always remain positive.

\subsection{\label{subsec:bogs=0}Scalar modes}
First, consider the Hawking radiation from massless scalar mode excitations, with spin ${s=0}$. The more general bosonic cases (${s=1}$ for photons and ${s=2}$ for gravitons) will be considered in the next section.

\begin{figure*}[t]
\centering
\begin{minipage}[l]{0.88\columnwidth}
  \includegraphics[width=\columnwidth]{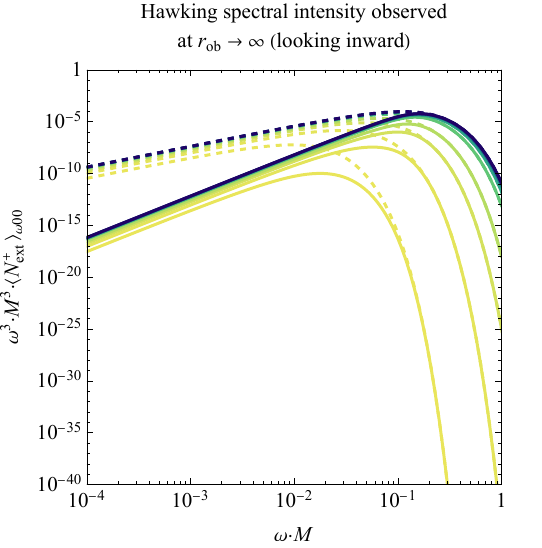}
\end{minipage}%
\hspace{1.5em}
\begin{minipage}[r]{0.88\columnwidth}
    \includegraphics[width=\columnwidth]{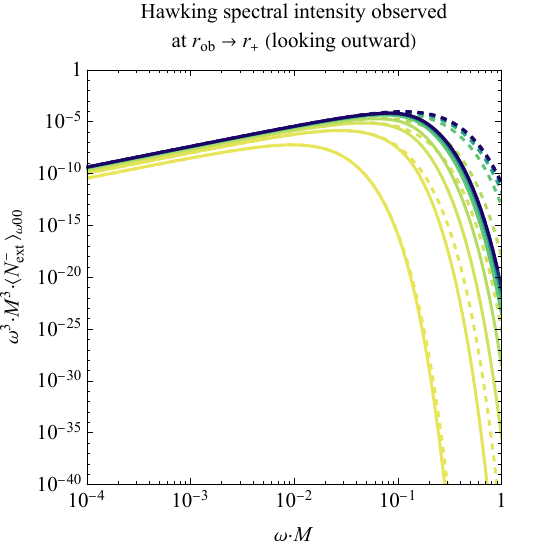}
\end{minipage}%
\vspace{1em}
\begin{minipage}[l]{0.88\columnwidth}
  \includegraphics[width=\columnwidth]{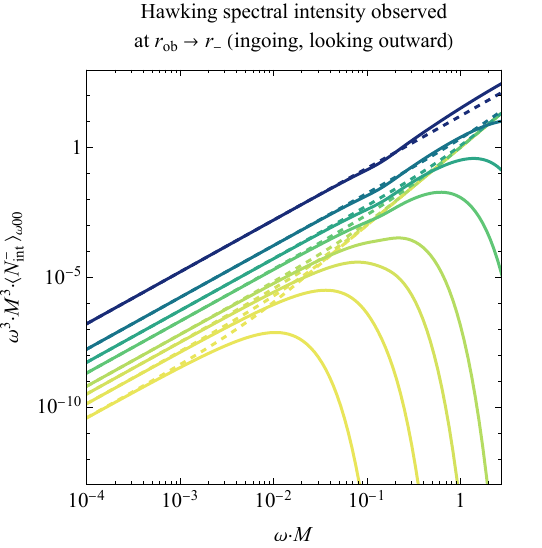}
\end{minipage}%
\hspace{1.5em}
\begin{minipage}[r]{0.88\columnwidth}
    \includegraphics[width=\columnwidth]{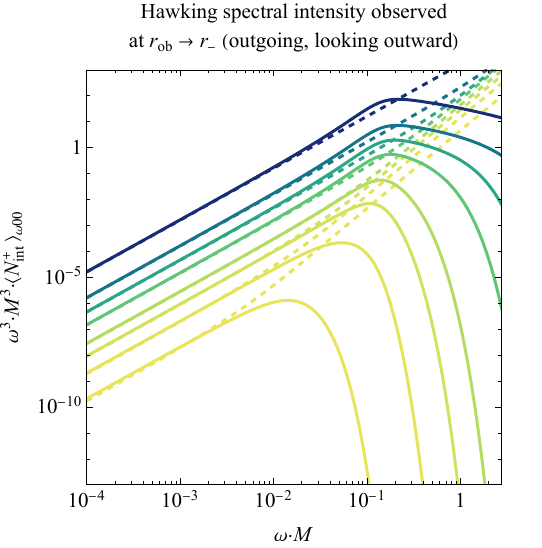}
\end{minipage}
\caption{Graybody spectra for the Hawking $s$-modes seen by an infalling observer at infinity (upper left panel), the event horizon (upper right panel), and the Cauchy horizon (lower left panel for an ingoing observer and lower right panel for an outgoing observer) in the Kerr spacetime with black hole spins ${a/M}$ = 0.1, 0.3, 0.5, 0.7, 0.9, 0.96, 0.99, and 0.999 (labeled with respective colors from dark blue to yellow). Dashed curves show the corresponding positive-valued blackbody spectra with temperatures $\varkappa_+/(2\pi)$ (upper two panels) and $\varkappa_-/(2\pi)$ (lower two panels) from Eq.~(\ref{eq:surf}), while the solid curves are evaluated numerically from Eqs.~(\ref{eq:N}).\label{fig:N00}}
\end{figure*}

The Hawking spectra for the lowest set of modes (${\ell=0,\ m=0}$) are shown in Fig.~\ref{fig:N00}. These $s$-wave spectra are computed numerically for a variety of black hole spin parameters seen by the four observers represented in Eqs.~(\ref{eq:N}). First, the standard graybody spectrum seen asymptotically far from the black hole is shown in the upper left panel of Fig.~\ref{fig:N00}. The distribution is plotted as a spectral intensity, which scales as ${\omega^3\langle N\rangle}$, so that a Planckian blackbody would appear with a quadratic power law at low frequencies and an exponential drop at high frequencies. Such a blackbody, with a temperature given by the surface gravity ${\varkappa_+/(2\pi)}$, is plotted for each spin parameter with a dashed curve. All the numerically-evaluated solid curves agree with the blackbody estimations at high frequencies (i.e., the geometric optics limit). However, at low frequencies, the graybody spectra differ from their blackbody counterparts by a power law index of 2, in agreement with the analytic prediction of Starobinsky in the limit ${\omega\to0}$ \cite{sta73}.

While the spectrum of Hawking radiation seen by someone looking inward from asymptotically far away contains entirely straightforward graybody deviations at low frequencies, the spectrum seen by someone crossing the event horizon contains graybody deviations at high frequencies, as shown in the upper right panel of Fig.~\ref{fig:N00}. The spectral intensity is still roughly the same order of magnitude as that seen at infinity, in accordance with assertion $\#$1 in Sec.~\ref{sec:int}. But at higher frequencies, the spectrum drops to zero much more quickly than one might expect from blackbody predictions. The reason for the dropoff is that in the geometric optics limit, fewer outgoing modes originating from an Unruh emitter at the horizon will be reflected and return to the observer; instead, more will escape as rays to infinity instead of being back-scattered as waves. Since the Hawking spectrum seen by an observer looking outward from the event horizon is determined entirely by these reflected modes (and not from transmitted ingoing modes originating from past null infinity, which do not exhibit the characteristic exponential peeling), Hawking radiation detected at the event horizon is suppressed at high frequencies.

Before discussing the Hawking spectrum seen by someone at the inner horizon, one feature present in all panels of Fig.~\ref{fig:N00} is the suppression of Hawking radiation for faster-spinning black holes. As the spin parameter $a$ is increased and the curves change color from dark blue to yellow, one may note that the higher-$a$ curves have overall lower intensities. The faster a black hole spins, the colder it becomes, regardless of where the observer lies within the spacetime.

The Hawking spectra seen by an observer at the Cauchy horizon are shown in the lower two panels of Fig.~\ref{fig:N00}. The lower left panel corresponds to an ingoing observer looking outward at the sky above (the left portion of $r_-$ in Fig.~\ref{fig:penrose}), while the lower right panel corresponds to an outgoing observer looking inward at the past horizon below (the right portion of $r_-$ in Fig.~\ref{fig:penrose}). These spectra are plotted alongside the dashed blackbody curves (after taking the absolute value) for the negative temperature given by the surface gravity of the inner horizon, $\varkappa_-/(2\pi)$. At low frequencies, all the curves approach the same quadratic power law, but instead of simply falling off exponentially as the frequency $\omega$ is increased, the curves continue to climb orders of magnitude higher than any positive-temperature blackbody would allow. However, the physically measurable spectral intensity does not contain an ultraviolet divergence for non-zero rotation. Eventually, as suspected in the Reissner-Nordstr\"om case \cite{mcm23a}, the exponential Wien tail dominates as ${\omega\to\infty}$ so that the spectral intensity returns to zero. But as the spin $a$ decreases and the inner horizon approaches the ${r=0}$ singularity, Hawking radiation is able to access higher and higher frequencies before reaching an ultraviolet cutoff. The Schwarzschild limit ${a/M\to0}$ is not shown in these lower two panels, since the spectral intensity in that case becomes infinite at all frequencies.

\begin{figure*}[t]
\centering
\begin{minipage}[l]{0.9\columnwidth}
  \includegraphics[width=\columnwidth]{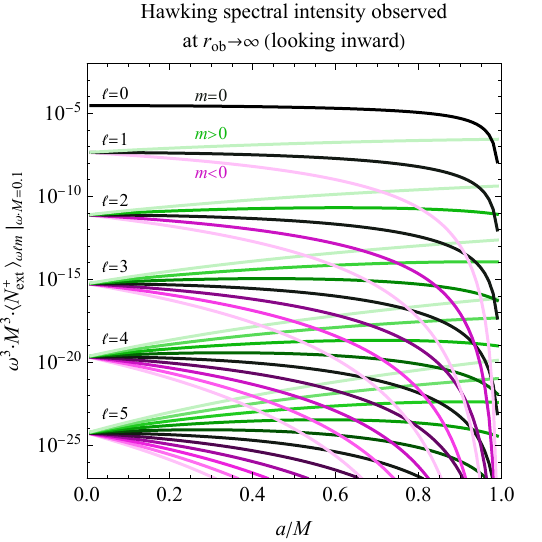}
\end{minipage}%
\hspace{1.5em}
\begin{minipage}[r]{0.9\columnwidth}
    \includegraphics[width=\columnwidth]{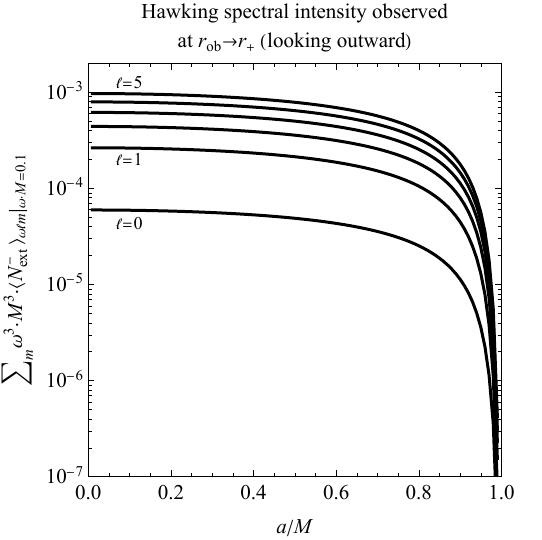}
\end{minipage}%
\vspace{1em}
\begin{minipage}[l]{0.9\columnwidth}
  \includegraphics[width=\columnwidth]{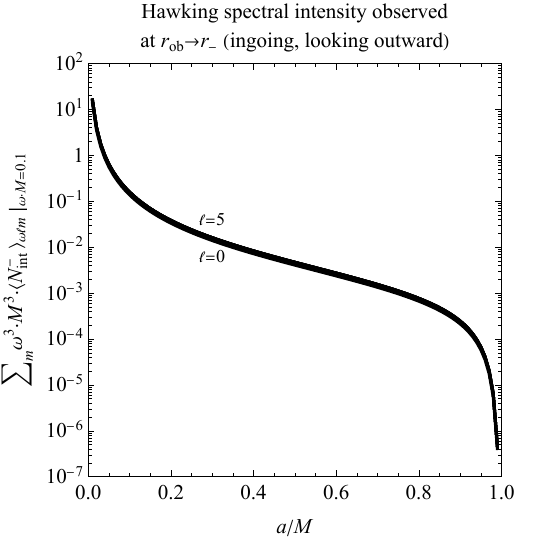}
\end{minipage}%
\hspace{1.5em}
\begin{minipage}[r]{0.9\columnwidth}
    \includegraphics[width=\columnwidth]{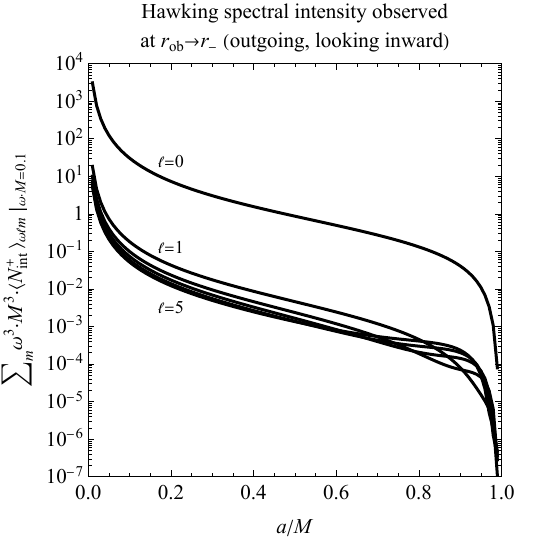}
\end{minipage}
\caption{Hawking spectral intensities for higher $\ell$- and $m$-modes, seen by an infalling observer at infinity (upper left panel), the event horizon (upper right panel), and the Cauchy horizon (lower two panels) in the Kerr spacetime as a function of the black hole spin parameter ${a/M}$. All modes are evaluated at a frequency of ${\omega\cdot M=0.1}$. Each individual $m$-mode for a given $\ell$ are presented for the observer at infinity, while the $m$-modes are summed for each $\ell$ for the other three observers.\label{fig:Nlm}}
\end{figure*}

It should be noted that the Cauchy horizon Hawking spectra shown in Fig.~\ref{fig:N00} are not \emph{a priori} expected to diverge. The divergent negative Hawking temperatures seen at the Cauchy horizon correspond to ingoing observers looking inward ($\kappa^+$ for ${P_r>0}$) and outgoing observers looking outward ($\kappa^-$ for ${P_r<0}$). In contrast, Fig.~\ref{fig:N00} shows observers at the Cauchy horizon looking in the direction opposite the Penrose blueshift singularity\textemdash ingoing observers looking outward and outgoing observers looking inward. Calculations for the former two scenarios would involve the inner product of the Unruh emitter's Kruskal modes with the observer's Fourier-decomposed and back-propagated Kruskal modes \cite{lan18}, which is nonetheless expected to yield an infinite spectral intensity at all frequencies. What Fig.~\ref{fig:N00} shows is that in addition to the classical (and likely also semiclassical) blueshift singularity, an observer at the Cauchy horizon will see the entire sky around them glowing brightly with Hawking radiation.

Beyond the $s$-wave approximation, the Hawking spectral intensities for higher-$\ell$ modes are shown in Fig.~\ref{fig:Nlm}. Instead of showing entire spectra as functions of the frequency $\omega$, these spectral intensities are evaluated at a specific mid-range frequency (${\omega\cdot M=0.1}$) so that the dependence on the spin parameter $a$ can be plotted more clearly.

For an observer asymptotically far from the black hole, the higher-$\ell$ spectral intensities are shown in the upper left panel of Fig.~\ref{fig:Nlm}. As the black hole spin $a$ is taken to zero, all azimuthal $m$-modes within a given $\ell$ mode converge to the same value, as expected. These intensities in the low-$a$ limit drop off as $\ell$ increases, so that the lowest angular mode dominates the Hawking spectrum. In particular, the $\ell$-modes are spaced apart by three to five orders of magnitude, in agreement with the approximate behavior predicted by Starobinsky \cite{sta73}:
\begin{equation}
    \left|\mathcal{T}^+_\text{ext}\right|^2\propto\frac{(\omega\cdot M)^{2\ell+1}(\ell!)^4}{[(2\ell)!]^2[(2\ell+1)!!]^2}.
\end{equation}
Additionally, note that in the upper left panel of Fig.~\ref{fig:Nlm}, the green curves (corresponding to positive $m$, with ${m=\ell}$ located highest in each group) always lie above the corresponding ${m=0}$ curves, while the magenta curves (corresponding to negative $m$, with ${m=-\ell}$ located lowest in each group) always lie below the corresponding ${m=0}$ curves. Physically, Hawking particles are always being emitted preferentially with the same angular momentum as the black hole, so that over time, the black hole will tend to spin down as the Hawking particles carry away excess angular momentum \cite{haw75}.

In the upper right panel of Fig.~\ref{fig:Nlm}, the Hawking spectral intensity is shown for an observer crossing the event horizon, yielding quite different behavior than that of an observer far away. While the ${\ell=0}$ mode dominates for an asymptotically distant observer, all higher-$\ell$ modes are present when the observer is in a regime where they are close enough to access more angular information than $s$-waves. While this panel plots the sum over all $m$-modes in a given $\ell$-mode, it should be noted that as ${\ell\to\infty}$, all the modes with ${m=0}$ tend to a constant value, just as quickly as all the modes with ${m=0}$ for an observer at infinity tend to zero\textemdash note the relationship between $\left|\mathcal{T}^+_{\text{ext},\omega}\right|^2$ and $\left|\mathcal{R}^-_{\text{ext},\omega}\right|^2$ in Eq.~(\ref{eq:norm}).

The lower two panels of Fig.~\ref{fig:Nlm} similarly show equal contributions from all higher-$\ell$ modes in the full Hawking spectrum seen by an infalling observer near the Cauchy horizon. Just as expected, the spectral intensity decreases as the black hole spin $a$ increases, and the intensity increases or decreases monotonically with $\ell$, except in the near-extremal case for an outgoing observer looking inward. The higher-$\ell$ behavior demonstrated in Fig.~\ref{fig:Nlm} matches that of the Reissner-Nordstr\"om model \cite{mcm23a}. Note that as ${a\to0}$, the observed Hawking spectral intensity diverges as the inner horizon meets the ${r=0}$ singularity, and as ${a\to M}$, the observed Hawking radiation vanishes as the inner horizon meets the outer horizon to create an extremal, zero-temperature spacetime.

\subsection{\label{subsec:bogs>0}Higher-spin modes}

\begin{figure}[t]
\centering
\includegraphics[width=\columnwidth]{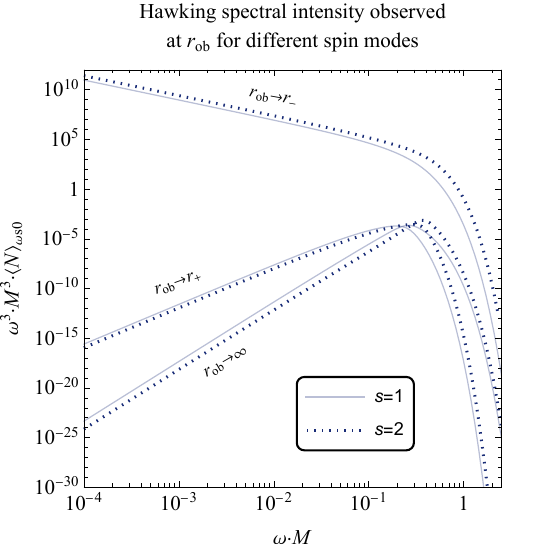}
\caption{Graybody spectra for the electromagnetic [solid curves, $(s,\ell,m)=(1,1,0)$] and gravitational [dotted curves, $(s,\ell,m)=(2,2,0)$] components of the Hawking radiation seen by an inertial observer at Boyer-Lindquist radius $r_\text{ob}$. All spectra are evaluated for a Kerr black hole with angular momentum ${a/M=0.1}$. These higher-spin spectra are qualitatively similar to their scalar counterparts from Fig.~\ref{fig:N00}, except in the case of an outgoing observer at the inner horizon, who sees an infrared-divergent spectrum.\label{fig:Ns}}
\end{figure}

While scalar modes (with spin 0) are commonplace in calculations of semiclassical effects in curved spacetimes due to the scalar wave equation's simplicity and the ``physical enough'' interpretation of modeling a single degree of freedom from a photon field, one can obtain more physically meaningful results by considering the higher-spin generalization of the wave equation given by Eq.~(\ref{eq:waveeq_r_R}).

The Hawking particles that will be considered here are photons from an electromagnetic field (spin-1) and gravitons from a gravitational field (spin-2). For all integer spins, the spectra of Eqs.~(\ref{eq:N}) remain Planckian, with the only modifications arising from the scattering coefficients calculated from each spin's corresponding wave equation \cite{pag76}.

The spectra for Hawking radiation from the lowest $\ell$- and $m$-modes of spin-1 and spin-2 fields are shown in Fig.~\ref{fig:Ns}. In this plot, it can be seen that the higher-spin fields radiate with roughly the same spectra as in the scalar case of Fig.~\ref{fig:N00} for exterior observers. When ${r_\text{ob}\to\infty}$, the spin-1 and spin-2 spectra peak at a slightly higher frequency than the spin-0 spectrum (not shown), but they also trail off in the infrared regime with a steeper power-law slope than the spin-0 spectrum. Similarly, the spectra for an observer at the event horizon appear almost identical for different values of spin.

All the spectra in Fig.~\ref{fig:Ns} are shown for a black hole with angular momentum ${a/M=0.1}$. Other values of $a$ yield qualitatively similar results (just as in Fig.~\ref{fig:N00}), but the low value of $a$ here is chosen since it gives the most pronounced effects, especially given the proximity of the inner horizon to the central singularity as ${a\to0}$.

For an observer at the inner horizon, only the spectrum seen by an outgoing observer looking inward is shown, since the spectra seen by an ingoing observer there for positive values of the spin weight $s$ are suppressed by the vanishing of the interior reflection coefficients dictated by the Heaviside function in Eq.~(\ref{eq:Bref-}). If instead one chooses ${s=-1}$ and ${s=-2}$ (i.e., the outgoing radiative parts of the field; see Footnote~\ref{foot:spinweight}), as is common in Kerr perturbation calculations for numerical feasibility, the inner horizon spectra will appear even more ultraviolet-divergent than in the scalar case. But even for the positive spin weights shown in Fig.~\ref{fig:Ns}, the spectrum of radiation produced from electromagnetic and gravitational modes is infrared-divergent, indicating that an outgoing observer looking inward at the exponentially redshifting and dimming surface of the start that collapsed long ago will see that surface glow more and more brightly in the infrared as they approach the Cauchy horizon.

\section{\label{sec:dis}Discussion}

Hawking radiation is an observer-dependent phenomenon. Here we have generalized the Bogoliubov coefficient calculation of Hawking \cite{haw74} to determine the effective temperature of semiclassical radiation seen in the vacuum state of the locally inertial rest frame of an arbitrary observer within the Kerr spacetime. Hawking found that if the observer is placed asymptotically far from the black hole, they will see a small amount of approximately thermal radiation emerge from the vacuum; here we explore the vast parameter space of various classes of observers both inside and outside of the black hole, all of whom will generally see a non-zero amount of Hawking radiation, sometimes thermal, sometimes not. This radiation appears to originate from the black hole shadow; i.e., the dimming, redshifting surface of the star that collapsed long ago to form the black hole.

The main goal of this study has been to extend the results of prior studies of Hawking radiation in Schwarzschild \cite{ham18} and Reissner-Nordstr\"om \cite{mcm23a} black holes to the more astrophysically relevant case of rotating Kerr black holes. While the prior studies benefited from spherical symmetry and needed only to examine freely falling radial observers at different radii $r$, Kerr black holes possess only azimuthal symmetry, so that the results differ also as an observer's latitude $\theta$, angular momentum $\mathcal{L}$, and Carter constant $\mathcal{K}$ are varied.

Here we have examined five classes of observers: those falling along the axis of rotation (Sec.~\ref{subsubsec:on-axis}); in a horizostationary orbit (Sec.~\ref{subsubsec:horizo}); falling along the equatorial plane with zero angular momentum (Sec.~\ref{subsec:ZAMO}); falling along the equatorial plane with zero energy (Sec.~\ref{subsec:CO}); and crossing through either the event horizon or the Cauchy horizon with arbitrary finite orbital parameters (Sec.~\ref{sec:bog}). In all cases, a non-zero amount of Hawking radiation is seen regardless of where the observer is.

More importantly, from an analysis of the effective temperature, whenever an observer reaches the Cauchy horizon, the Hawking radiation will become blueshift-divergent (i.e., negative-temperature) in at least one direction in their field of view. Such a divergence was not explicitly calculated in the spectral analysis of Sec.~\ref{sec:bog}; instead, the cases where the Eddington-Finkelstein modes lead to a 1D scattering problem demonstrate that an inner-horizon observer looking in the \emph{opposite} direction from the anticipated semiclassical divergence will still see a high-energy glow of Hawking radiation. It is worth noting that in a dynamical gravitational collapse, the left portion of the inner horizon reached by an ingoing observer may not be a true Cauchy horizon like it is in the Kerr metric \cite{kra06}. In such a case, any Hadamard state should remain singularity-free along that part of the inner horizon, in accordance with the Fulling-Sweeny-Wald theorem \cite{ful78}. However, the divergent behavior at the right portion of the inner horizon reached by an outgoing observer remains valid, and the authors nevertheless expect the conclusions of this study at all horizons to remain robust for any astronomically realistic black hole \cite{ham10}.

We additionally conclude here that the amount of Hawking radiation seen by different Kerr observers remains relatively independent of the spin of the quantum field, the particular $\ell m$-mode of the quantum state, and the direction in which the observer looks. Though all of these variables yield differing results for the Hawking radiation calculations, the key takeaways concerning the ubiquity of non-zero semiclassical particle creation and an inner horizon divergence remain unchanged.

One of the biggest outstanding questions that one may ask concerning this analysis of Hawking radiation in the framework of semiclassical gravity is how the radiation back-reacts on the spacetime. Here we have kept the spacetime geometry fixed, but presumably if enough radiation is present, the particles produced will possess a gravitational field of their own that can change the underlying metric. Usually, one assumes that these back-reaction effects are negligibly small and the Kerr metric can still be used as a valid approximation of the spacetime geometry for astrophysical, rotating black holes. We find this assumption to be generally true for observers outside of the event horizon. However, close to the inner horizon, the generally divergent behavior of observed Hawking radiation suggests that the Kerr metric is not semiclassically self-consistent there. The back-reaction for astrophysically relevant perturbations is likely to form a strong, spacelike singularity just above the inner horizon \cite{mcm21}, though other mathematical predictions also exist in the literature, such as a weak, null singularity \cite{daf05} or a rapid implosion toward the formation of a compact horizonless object \cite{bar21}.

The obvious problem with attempting to analyze the effects of back-reaction from Hawking radiation in the present framework is that each observer sees a different amount of Hawking radiation, but they all exist in the same background metric. Even though the radiation is completely real from the perspective of any individual observer, the very definition of a particle depends on an observer's frame of reference, a concept which seems at first glance completely at odds with the central claim in semiclassical gravity that particles arising from quantum fields provide the source of curvature for a global, classical spacetime metric.

Which observer is the one to see the actual radiation that contributes to the underlying spacetime geometry? Is there a preferred quantum reference frame, or will each observer construct their own background based on the particular back-reaction they see in their own frame? The usual approach in semiclassical gravity is to place a quantum field in a particular global state and construct an averaged version of that field's net energy-momentum, which can be cast into a Lorentz-covariant (and observer-independent) form \cite{mol59,ros63}. This quantity, the renormalized vacuum expectation value ${\langle T^{\mu\nu}\rangle_\text{ren}}$ of the field's stress-energy tensor (RSET), is then guaranteed to feed back into the spacetime geometry in the standard classical way.

However, the RSET is only an averaged quantity and cannot provide all the information that an observer is able to access about the state of a quantum field in a curved spacetime. Classically, the stress-energy tensor involves an integral with each particle's energy-momentum over a Lorentz-invariant pseudo-scalar volume element:
\begin{equation}
    T^{\mu\nu}=\int p^\mu p^\nu f(\bm{x},\bm{p})\frac{g\ d^3p}{p^0(2\pi\hbar)^3},
\end{equation}
where $g$ is the number of spin states of the particle and $f$ is the dimensionless occupation number, which specifies the number $dN$ of particles with 4-position $x^\mu$ and 4-momentum $p^\mu$ within the Lorentz-invariant six-dimensional volume of phase space ${d^3x\ d^3p}$. The particle number $N$, directly related to this occupation number $f$, is precisely the object of study throughout this work. While the RSET is a difficult object to calculate even for the most symmetric spacetimes, one of the goals of this work is to show that ${\langle N\rangle}$, the more elementary object, is entirely straightforward to calculate in the locally inertial frame of any observer.

In conclusion, by examining the expectation value of the number operator for a quantum field placed over a Kerr spacetime, we have analyzed and extended the same key ideas that were anticipated in Sec.~\ref{sec:int} from prior studies of spherically symmetric black holes: (1) The Hawking radiation seen by generic observers passing the vicinity of the event horizon has a negligibly weak graybody spectrum, (2) the Hawking radiation seen at the Cauchy horizon possesses a divergent effective temperature for all the classes of observers examined, (3) the Hawking radiation originating from different directions in the sky varies considerably for different classes of observers, and (4) the effective Hawking temperature for certain observers can become negative even outside of the event horizon, though an ultraviolet divergence in the Hawking spectrum is not seen for the limited cases considered here. From these results it is clear that the Kerr metric cannot be trusted in its full global form in the semiclassical approximation, as a result of the diverging quantum radiation that would be observed at the Cauchy horizon.

\appendix
\section{\label{app:mst}Interior mode scattering in the MST method}

The Black Hole Perturbation Toolkit (BHPT) \cite{bhp} makes use of the Mano-Suzuki-Takasugi (MST) method \cite{man96,man97} to calculate the scattering coefficients for a Klein-Gordon field with spin weight $s$ in the Kerr spacetime. In this Appendix, we review the implementation of this method for the scattering of exterior modes between the event horizon and spatial infinity, and we extend the analysis to include the scattering of interior modes between the event horizon and the Cauchy horizon.

The notation used throughout this Appendix is self-consistent but may differ from the notation used in the main body of the paper; instead it is chosen to match that of the BHPT and its relevant references. The most notable change is that $R$ here no longer represents the scale length defined below Eq.~(\ref{eq:kerr}) and instead represents the Teukolsky radial mode function defined via the mode expansion
\begin{equation}\label{eq:modesep_R}
    \phi_{\omega\ell m}=\frac{R_{\omega\ell m}(r)\ {}_sS^\omega_{\ell m}(\theta)\ \text{e}^{im\varphi-i\omega t}}{\sqrt{4\pi\omega}},
\end{equation}
vis-\`a-vis Eq.~(\ref{eq:modesep}) (therefore ${\psi_{\omega\ell}(r)=R_{\omega\ell m}(r)\sqrt{r^2+a^2}}$). The modes ${{}_sS^\omega_{\ell m}(\theta)}$ represent spin-weighted spheroidal wave functions \cite{ber06}, while the radial modes ${R_{\omega\ell m}(r)}$ satisfy the homogeneous radial Teukolsky equation \cite{teu72}:
\begin{align}
    &\left(\frac{K_{\omega m}^2-2is(r-M)K_{\omega m}}{\Delta}+4is\omega r-\lambda_{\omega\ell m}\right)R_{\omega\ell m}\nonumber\\
    &+\Delta^{-s}\frac{d}{dr}\left(\Delta^{s+1}\frac{dR_{\omega\ell m}}{dr}\right)=0,\label{eq:waveeq_r_R}
\end{align}
for a black hole of mass $M$ and spin $a$, vis-\`a-vis Eq.~(\ref{eq:waveeq_r}). The horizon function $\Delta$ is defined in the same way as in the text following Eq.~(\ref{eq:kerr}), the function ${K_{\omega m}\equiv(r^2+a^2)\omega-ma}$, and the constant ${\lambda_{\omega\ell m}\equiv\mathscr{E}_{\ell m}-2ma\omega+a^2\omega^2-s(s+1)}$, where $\mathscr{E}_{\ell m}$ is the eigenvalue of the spin-weighted spheroidal wave function ${{}_sS^{\omega}_{\ell m}(\theta)}$ \cite{fle59} and reduces to ${\mathscr{E}_{\ell m}\to\ell(\ell+1)}$ in the Schwarzschild limit.

The BHPT allows for the following boundary value problem to be solved: consider two sets of modes $R_{\omega\ell m}^{\text{in}}$ (initialized on past null infinity) and $R_{\omega\ell m}^{\text{up}}$ (initialized on the past horizon), which asymptotically approach the boundary values
\begin{equation}\label{eq:R_in_asymptotic}
    R^{\text{in}}_{\omega\ell m}\to
    \begin{cases}
        B^{\text{ref}}_{\text{ext}}r^{-1-2s}\text{e}^{i\omega r^*}+B^{\text{inc}}_{\text{ext}}r^{-1}\text{e}^{-i\omega r^*},&r\to\infty\\
        B^{\text{trans}}_{\text{ext}}|\Delta|^{-s}\text{e}^{-i\omega_+r^*},&r\to r_+\\
        B^{\text{ref}}_{\text{int}}\ \text{e}^{i\omega_-r^*}+B^{\text{trans}}_{\text{int}}|\Delta|^{-s}\text{e}^{-i\omega_-r^*},&r\to r_-
    \end{cases},
\end{equation}
\begin{equation}\label{eq:R_up_asymptotic}
    R^{\text{up}}_{\omega\ell m}\to
    \begin{cases}
        C^{\text{trans}}_{\text{ext}}r^{-1-2s}\text{e}^{i\omega r^*},&r\to\infty\\
        C^{\text{up}}_{\text{ext}}\ \text{e}^{i\omega_+r^*}+C^{\text{ref}}_{\text{ext}}|\Delta|^{-s}\text{e}^{-i\omega_+r^*},&r\to r_+\\
        C^{\text{trans}}_{\text{int}}\ \text{e}^{i\omega_-r^*}+C^{\text{ref}}_{\text{int}}|\Delta|^{-s}\text{e}^{-i\omega_-r^*},&r\to r_-
    \end{cases},
\end{equation}
where $\omega_\pm$ is given by Eq.~(\ref{eq:omega_pm}) and the complex constants $B$ and $C$ are scattering coefficients for either the reflection, incidence, or transmission of the mode waves. The tortoise coordinate $r^*$ is chosen to be
\begin{equation}
    r^*\equiv r+\frac{1}{2\varkappa_+}\ln\left|\frac{r-r_+}{2M}\right|+\frac{1}{2\varkappa_-}\ln\left|\frac{r-r_-}{2M}\right|,
\end{equation}
with the surface gravity $\varkappa_\pm$ (which is negative at the inner horizon) given by Eq.~(\ref{eq:surf}). Note that the forward-propagated modes $R_{\omega\ell m}$ and scattering coefficients $B$ and $C$ defined here are different from the backward-propagated modes $\psi_\text{ob}$ and scattering coefficients $\mathcal{T}$ and $\mathcal{R}$ used in the main text; the relation between the two will be given at the end of this Appendix.

Conservation of the Wronskian in the scalar case leads to the following relations between the scattering coefficients:
\begin{align}
    &\left|B^{\text{ref}}_{\text{ext}}\right|^2+\frac{\omega_+(r_+^2+a^2)}{\omega}\left|B^{\text{trans}}_{\text{ext}}\right|^2=\left|B^{\text{inc}}_{\text{ext}}\right|^2,\nonumber\\
    &\left|C^{\text{ref}}_{\text{ext}}\right|^2+\frac{\omega}{\omega_+(r_+^2+a^2)}\left|C^{\text{trans}}_{\text{ext}}\right|^2=\left|C^{\text{up}}_{\text{ext}}\right|^2,\nonumber\\
    &\frac{C^{\text{trans}}_{\text{ext}}}{C^{\text{up}}_{\text{ext}}}=\frac{\omega_+(r_+^2+a^2)}{\omega}\frac{B^{\text{trans}}_{\text{ext}}}{B^{\text{inc}}_{\text{ext}}},\nonumber\\
    &\frac{C^{\text{trans}}_{\text{ext}}}{C^{\text{ref}}_{\text{ext}}}=-\frac{\omega_+(r_+^2+a^2)}{\omega}\frac{B^{\text{trans*}}_{\text{ext}}}{B^{\text{ref*}}_{\text{ext}}},\nonumber\\
    &\left|B^{\text{ref}}_{\text{int}}\right|^2+\frac{\omega_+(r_+^2+a^2)}{\omega_-(r_-^2+a^2)}\left|B^{\text{trans}}_{\text{ext}}\right|^2=\left|B^{\text{trans}}_{\text{int}}\right|^2,\nonumber\\
    &\left|C^{\text{up}}_{\text{int}}\right|^2-\frac{\omega_-(r_-^2+a^2)}{\omega_+(r_+^2+a^2)}\left(\left|C^{\text{trans}}_{\text{int}}\right|^2-\left|C^{\text{ref}}_{\text{int}}\right|^2\right)=\left|C^{\text{ref}}_{\text{ext}}\right|^2,\nonumber\\
    &B^{\text{trans}}_{\text{int}}C^{\text{trans}}_{\text{int}}-B^{\text{ref}}_{\text{int}}C^{\text{ref}}_{\text{int}}=\frac{\omega_+(r_+^2+a^2)}{\omega_-(r_-^2+a^2)}B^{\text{trans}}_{\text{ext}}C^{\text{up}}_{\text{int}},\nonumber\\
    &B^{\text{trans*}}_{\text{int}}C^{\text{ref}}_{\text{int}}- B^{\text{ref*}}_{\text{int}}C^{\text{trans}}_{\text{int}}=\frac{\omega_+(r_+^2+a^2)}{\omega_-(r_-^2+a^2)}B^{\text{trans*}}_{\text{ext}}C^{\text{ref}}_{\text{ext}},
\end{align}
where a superscript asterisk ($^*$) here and elsewhere denotes complex conjugation, except in the case of the tortoise coordinate $r^*$. In what follows, we will focus on the \emph{in} modes; the scattering coefficients for the \emph{up} modes can be obtained from the \emph{in} modes through the above Wronskian conditions. The main results for the application of the MST method in the exterior portion of the spacetime will be quoted here; for a more complete review, see Ref.~\cite{sas03}.

The solutions to the radial wave Eq.~(\ref{eq:waveeq_r_R}) belong to a class of functions known as confluent Heun functions. However, the mathematical properties of these functions (especially their asymptotic behavior at each horizon) has been a mathematical enigma, and to this day, the central two-point connection problem for these functions still has no explicit solution. However, Svartholm \cite{sva39} and Erd{\'e}lyi \cite{erd42,erd44} early on discovered an integral transform for confluent Heun functions with a hypergeometric kernel, so that they could be expressed in a series representation as the sum of (the more tractable) hypergeometric functions.

Under the MST method, two infinite series expansions of the solutions to Eq.~(\ref{eq:waveeq_r_R}) are found that are valid in different but overlapping regimes. The first of these expansions, in terms of ordinary hypergeometric functions ${{}_2F_1}$, is valid for all finite values of $r$ but breaks down as ${r\to\infty}$. Defining the dimensionless parameters\footnote{As a reminder, the parameters defined here and used throughout Appendix~\ref{app:mst} should be treated independently from the notation of the main body of the paper and instead are chosen to align with the notation used by MST \cite{man96,man97}. In particular, $\kappa$ does not represent the effective temperature, $\epsilon$ does not represent the adiabatic control function, and $\tau$ does not represent proper time.}
\begin{gather}
    x\equiv\frac{\omega}{\epsilon\kappa}(r_+-r),\quad\epsilon\equiv2M\omega,\quad\kappa\equiv\sqrt{1-\left(\frac{a}{M}\right)^2},\nonumber\\
    \epsilon_\pm\equiv\frac{\epsilon\pm\tau}{2},\quad\tau\equiv\frac{\epsilon-m\left(\frac{a}{M}\right)}{\kappa},\label{eq:MST_parameters}
\end{gather}
such that the outer ($+$) and inner ($-$) horizon radii are given by ${r_\pm\equiv(1\pm\kappa)M}$, this first series is:
\begin{align}\label{eq:R_hypergeometric}
    R^{\text{in}}_{\nu}&=\text{e}^{i\epsilon\kappa x}|-x|^{-s-i\epsilon_+}|1-x|^{i\epsilon_-}\sum_{n=-\infty}^{\infty}a_n^\nu(s)\nonumber\\
    &\times{}_2F_1(n+\nu+1-i\tau,-n-\nu-i\tau;1-s-2i\epsilon_+;x),
\end{align}
where the parameter $\nu$, called the renormalized angular momentum, is a generalization of $\ell$ to non-integer values that is fixed by requiring that the series solution to the Teukolsky equation converges.

Likewise, the second series expansion can be written in terms of confluent hypergeometric functions and is valid for asymptotically large values of $r$ but fails as ${r\to r_+}$. By matching these two expansions, the coefficients $a_n^\nu(s)$ in both expansions will satisfy the same three-term recurrence relation that can be solved numerically to find the minimal solution.

As ${r\to r_+}$ (or equivalently, as ${x\to 0}$), Eq.~(\ref{eq:R_hypergeometric}) reduces to
\begin{equation}\label{eq:R_hypergeometric_+}
    R^{\text{in}}_{\nu}\to|-x|^{-s-i\epsilon_+}\sum_{n=-\infty}^{\infty}a_n^\nu(s),
\end{equation}
while Eq.~(\ref{eq:R_in_asymptotic}) can be written in terms of the parameters of Eq.~(\ref{eq:MST_parameters}) as
\begin{equation}\label{eq:R_in_asymptotic_+}
    R^{\text{in}}_{\omega\ell m}\to B^{\text{trans}}_{\text{ext}}\left(\frac{\epsilon\kappa}{\omega}\right)^{-2s}|-x|^{-s}\text{ e}^{-i\epsilon_+\left(\ln|-x|+\kappa+\frac{2\kappa\ln\kappa}{1+\kappa}\right)}.
\end{equation}
The coefficient $B^{\text{trans}}_{\text{ext}}$ can then be read off by equating the two expressions from Eqs.~(\ref{eq:R_hypergeometric_+}) and (\ref{eq:R_in_asymptotic_+}). A similar matching process leads to expressions for the scattering parameters at infinity. The resulting formulae for $B^{\text{trans}}_{\text{ext}}$, $B^{\text{inc}}_{\text{ext}}$, and $B^{\text{ref}}_{\text{ext}}$ are given respectively by Eqs.~(167)\textendash(169) of Ref.~\cite{sas03}.

Now, consider how the above formalism may be extended to the black hole's interior. As ${r\to r_-}$ (or equivalently, as ${x\to1}$), based on the asymptotic behavior of the hypergeometric function with argument unity, Eqs.~(\ref{eq:R_hypergeometric}) and (\ref{eq:R_in_asymptotic}) respectively reduce to
\begin{widetext}
\begin{align}
    R^{\text{in}}_\nu\to\text{e}^{i\epsilon\kappa}&\Bigg[|1-x|^{i\epsilon_-}\sum_{n=-\infty}^{\infty}a_n^\nu(s)\frac{\Gamma(1-s-2i\epsilon_+)\Gamma(-s-2i\epsilon_-)}{\Gamma(-n-\nu-s-i\epsilon)\Gamma(n+\nu+1-s-i\epsilon)}H(-s)\nonumber\\
    &+|1-x|^{-s-i\epsilon_-}\sum_{n=-\infty}^{\infty}a_n^\nu(s)\frac{\Gamma(1-s-2i\epsilon_+)\Gamma(s+2i\epsilon_-)}{\Gamma(-n-\nu-i\tau)\Gamma(n+\nu+1-i\tau)}H(s)\Bigg],\label{eq:R_hypergeometric_-}
\end{align}
\begin{equation}\label{eq:R_in_asymptotic_-}
    R^{\text{in}}_{\omega\ell m}\to B^{\text{ref}}_{\text{int}}\text{ exp}\left[i\epsilon_-\left(\ln|1-x|-\kappa-\frac{2\kappa\ln\kappa}{1-\kappa}\right)\right]+B^{\text{trans}}_{\text{int}}\left(\frac{\epsilon\kappa}{\omega}\right)^{-2s}|1-x|^{-s}\text{ exp}\left[-i\epsilon_-\left(\ln|1-x|-\kappa-\frac{2\kappa\ln\kappa}{1-\kappa}\right)\right],
\end{equation}
\end{widetext}
where ${H(s)}$ is the Heaviside step function defined by
\begin{equation}
    H(s)\equiv
    \begin{cases}
        1,&s\geq0\\
        0,&s<0
    \end{cases}.
\end{equation}
Eq.~(\ref{eq:R_hypergeometric_-}) breaks down for scalar modes when either ${\omega=0}$ or ${a=0}$; these cases, whose modes asymptotically scale as ${\ln(1-x)}$, must be treated separately.

Matching Eqs.~(\ref{eq:R_hypergeometric_-}) and (\ref{eq:R_in_asymptotic_-}) leads to expressions for the internal scattering parameters:
\begin{align}
    &B^{\text{ref}}_{\text{int}}=H(-s)\text{ e}^{i\kappa\left[\epsilon+\epsilon_-\left(1+\frac{2\ln\kappa}{1-\kappa}\right)\right]}\nonumber\\
    &\times\sum_{n=-\infty}^{\infty}a_n^\nu(s)\frac{\Gamma(1-s-2i\epsilon_+)\Gamma(-s-2i\epsilon_-)}{\Gamma(-n-\nu-s-i\epsilon)\Gamma(n+\nu+1-s-i\epsilon)},\label{eq:Bref-}
\end{align}
\begin{align}
    B^{\text{trans}}_{\text{int}}&=H(s)\left(\frac{\epsilon\kappa}{\omega}\right)^{2s}\text{ e}^{i\kappa\left[\epsilon-\epsilon_-\left(1+\frac{2\ln\kappa}{1-\kappa}\right)\right]}\nonumber\\
    &\times\sum_{n=-\infty}^{\infty}a_n^\nu(s)\frac{\Gamma(1-s-2i\epsilon_+)\Gamma(s+2i\epsilon_-)}{\Gamma(-n-\nu-i\tau)\Gamma(n+\nu+1-i\tau)}.\label{eq:Binc-}
\end{align}
The step functions in Eqs.~(\ref{eq:Bref-}) and (\ref{eq:Binc-}) imply that close to the inner horizon, only the non-negative (non-positive) spin-weighted components of ingoing (outgoing) waves survive, since these are the radiative (i.e., dominant propagating; see Footnote~\ref{foot:spinweight}) components.

The expressions above for the inner horizon scattering coefficients can then be implemented in Mathematica alongside the rest of the BHPT's Teukolsky package to compute the relevant \emph{in} scattering coefficients. In order to connect these results to the calculations in the main body of the paper, a transformation must be made between the two sets of scattering coefficients ($B$ and $C$ on the one hand and $\mathcal{T}$ and $\mathcal{R}$ on the other). One of the main problems that gives rise to the need for such a non-trivial matching is that the main text's back-propagated modes ${{}^{\text{ext,int}}\psi^\pm_\text{ob}}$ are initialized on the future null boundaries, while this Appendix's forward-propagated modes $R^{\text{in,up}}_{\omega\ell m}$ are initialized on the past null boundaries.


To find the back-propagated scattering coefficients, first note that through the conservation of the Wronskian, the transmission and reflection coefficients $\mathcal{T}^\pm_\text{ext,int}$ and $\mathcal{R}^\pm_\text{ext,int}$ must satisfy the normalization conditions
\begin{subequations}\label{eq:norm}
\begin{align}
    \left|\mathcal{R}^\pm_{\text{ext},\omega}\right|^2+\left|\mathcal{T}^\pm_{\text{ext},\omega}\right|^2\left(\frac{\omega\mp m\Omega_+}{\omega}\right)&=1,\\
    \left(\left|\mathcal{R}^\pm_{\text{int},\omega}\right|^2-\left|\mathcal{T}^\pm_{\text{int},\omega}\right|^2\right)\left(\frac{\omega_++m\Omega_-}{\omega}\right)&=1.
\end{align}
\end{subequations}

Consider first the exterior set of modes used to calculate $\langle N^\pm_\text{ext}\rangle_{\omega\ell m}$, encoded by a family of observers asymptotically close to future null infinity and the event horizon. Through time reversal, these modes ${{}^{\text{ext}}\psi_\text{ob}^\pm}$ map to the modes $R^{\text{in,up}}_{\omega\ell m}$ by the transformation ${(\omega,m)\mapsto(-\omega,-m)}$, which corresponds to taking the complex conjugate of the scattering coefficients, since
\begin{equation}
    R_{(-\omega)\ell(-m)}=R^*_{\omega\ell m},\qquad B^{\text{trans,ref}}_{(-\omega)\ell(-m)}=B^{\text{trans,ref*}}_{\omega\ell m}.
\end{equation}
By matching the asymptotic relations for the complete set of modes $R^{\text{in}}_{\omega\ell m}$ and $R^{\text{up}}_{\omega\ell m}$ in the spin-0 limit of Eqs.~(\ref{eq:R_in_asymptotic}) and (\ref{eq:R_up_asymptotic}) with the appropriately rescaled set of exterior modes ${{}^{\text{ext}}\psi^{-\text{*}}_\text{ob}}$ and ${{}^{\text{ext}}\psi^{+\text{*}}_\text{ob}}$ from Eqs.~(\ref{eq:f_obext+}) and (\ref{eq:f_obext-}), respectively, one arrives at the following equalities, after fixing ${B^\text{inc}_\text{ext}=1}$ and ${C^\text{up}_\text{ext}=1/\sqrt{r_+^2+a^2}}$:
\begin{subequations}\label{eq:scattermatch_ext}
\begin{align}
    &\mathcal{T}^+_{\text{ext},\omega}=B^{\text{trans*}}_{\text{ext},\omega}\sqrt{r_+^2+a^2},\\
    &\mathcal{R}^+_{\text{ext},\omega}=B^{\text{ref*}}_{\text{ext},\omega},\\
    &\mathcal{T}^-_{\text{ext},\omega_+}=C^{\text{trans*}}_{\text{ext},\omega},\\
    &\mathcal{R}^-_{\text{ext},\omega_+}=C^{\text{ref*}}_{\text{ext},\omega}\sqrt{r_+^2+a^2}.
\end{align}
\end{subequations}
Notice the addition of the mode frequency subscripts in the above scattering coefficients that help highlight a key difference between the ${(B,C)}$ coefficients and the ${(\mathcal{T},\mathcal{R})}$ coefficients\textemdash the scattering process defined by the former is initialized with the frequency eigenmodes of the wave equation, while the scattering process of the latter is initialized so that the observer always sees a frequency $\omega$.

Now consider the interior modes used to calculate $\langle N^\pm_\text{int}\rangle_{\omega\ell m}$, encoded by a family of observers reaching the ingoing and outgoing portions of the inner horizon (together with observers asymptotically close to future null infinity, to form a complete Cauchy slice). Since the global scattering process now depends on three singular points and the interior scattering potential is dynamical, the backward-propagated modes will not map trivially onto the forward-propagated modes by time reversal. One may instead consider the transformation ${(M,r)\mapsto(-M,-r)}$ as in Ref.~\cite{zil22b}, which leaves the radial wave Eq.~(\ref{eq:waveeq_r}) unchanged aside from a swapping of the asymptotic regimes at the inner and outer horizons. However, since that transformation leaves the irregular singular point at ${r\to\infty}$ unchanged, the forward- and backward-propagated modes must transform into linear combinations of one another.

In order to solve for the interior scattering coefficients, define coefficients $\alpha^\pm_\text{in,up}$ that form linear combinations of the modes $R^\text{in}_{\omega\ell m}$ and $R^\text{up}_{\omega\ell m}$. These linear combinations can be asymptotically equated to the modes ${{}^\text{int}\psi^-_\text{ob}}$ initialized along the ingoing portion of the future Cauchy slice and the modes ${{}^\text{int}\psi^{+\text{*}}_\text{ob}+{}^\text{ext}\psi^{+}_\text{ob}}$ initialized along the outgoing portions of the future Cauchy slice:
\begin{subequations}
\begin{align}
    \left(\alpha_\text{in}^-R^{\text{in}}_{\omega\ell m}+\alpha_\text{up}^-R^{\text{up}}_{\omega\ell m}\right)\sqrt{r^2+a^2}&={}^\text{int}\psi^{-}_\text{ob},\\
    \left(\alpha_\text{in}^+R^{\text{in}}_{\omega\ell m}+\alpha_\text{up}^+R^{\text{up}}_{\omega\ell m}\right)\sqrt{r^2+a^2}&={}^\text{int}\psi^{+\text{*}}_\text{ob}+{}^\text{ext}\psi^{+}_\text{ob}
\end{align}
\end{subequations}

Matching the asymptotic relations for the above sets of modes along the inner horizon leads to the linear coefficient values
\begin{subequations}
\begin{align}
    \alpha_\text{in}^+&=\frac{C^\text{ref}_\text{int}}{D},&&\alpha_\text{up}^+=-\frac{B^\text{trans}_\text{int}}{D},\\
    \alpha_\text{in}^-&=-\frac{C^\text{trans}_\text{int}}{D},&&\alpha_\text{up}^-=\frac{B^\text{ref}_\text{int}}{D},
\end{align}
\end{subequations}
where
\begin{equation}
    D\equiv\left(B^\text{ref}_\text{int}C^\text{ref}_\text{int}-B^\text{trans}_\text{int}C^\text{trans}_\text{int}\right)\sqrt{r_-^2+a^2}.
\end{equation}
The scattering coefficients can then be found by matching the asymptotic relations along the event horizon and at infinity:
\begin{subequations}\label{eq:scattermatch_int}
\begin{align}
    &\mathcal{T}^-_{\text{int},\omega_-}=\left(\alpha^{-}_{\text{in},\omega}B^\text{trans}_{\text{ext},\omega}+\alpha^{-}_{\text{up},\omega}C^\text{ref}_{\text{ext},\omega}\right)\sqrt{r_+^2+a^2},\\
    &\mathcal{R}^-_{\text{int},\omega_-}=\alpha^{-}_{\text{up},\omega}C^\text{up}_{\text{int},\omega}\sqrt{r_+^2+a^2},\displaybreak[0]\\
    &\mathcal{T}^+_{\text{int},\omega_-}=\alpha^{+\text{*}}_{\text{up},\omega}C^\text{up*}_{\text{int},\omega}\sqrt{r_+^2+a^2},\\
    &\mathcal{R}^+_{\text{int},\omega_-}=\left(\alpha^{+\text{*}}_{\text{in},\omega}B^\text{trans*}_{\text{ext},\omega}+\alpha^{+\text{*}}_{\text{up},\omega}C^\text{ref*}_{\text{ext},\omega}\right)\sqrt{r_+^2+a^2},\displaybreak[0]\\    
    &\mathcal{T}^-_{\text{ext},\omega_+}=\frac{\alpha^-_{\text{in},\omega}B^\text{inc}_{\text{ext},\omega}}{\mathcal{T}^-_{\text{int},\omega_-}},\\
    &\mathcal{R}^-_{\text{ext},\omega_+}=\frac{\alpha^-_{\text{up},\omega}C^\text{up}_{\text{ext},\omega}}{\mathcal{T}^-_{\text{int},\omega_-}}\sqrt{r_+^2+a^2},\displaybreak[0]\\
    &\mathcal{T}^+_{\text{ext},\omega}=\alpha^+_{\text{up},\omega}C^\text{up}_{\text{ext},\omega}\sqrt{r_+^2+a^2}-\mathcal{R}^{+\text{*}}_{\text{int},\omega_-}\mathcal{R}^{-\text{*}}_{\text{ext},\omega_+},\\
    &\mathcal{R}^+_{\text{ext},\omega}=\alpha^+_{\text{in},\omega}B^\text{inc}_{\text{ext},\omega}-\mathcal{R}^{+\text{*}}_{\text{int},\omega_-}\mathcal{T}^{-\text{*}}_{\text{ext},\omega_+}.
\end{align}
\end{subequations}
Note that, unlike in the exterior case of Eqs.~(\ref{eq:scattermatch_ext}), the $B^\text{inc}$ and $C^\text{up}$ coefficients from Eqs.~(\ref{eq:scattermatch_int}) do not need to be fixed since all the $B$ and $C$ coefficients are written in a normalization-free form. Thus, one may retain the default normalization choice ${B^\text{trans}=1}$ used by the BHPT.

Once the scattering coefficients $B$ and $C$ are computed with the help of the BHPT, Eqs.~(\ref{eq:scattermatch_ext}) can be used to calculate the back-scattering coefficients $\mathcal{T}$ and $\mathcal{R}$ used in Eqs.~(\ref{eq:N+ext})\textendash(\ref{eq:N-ext}), while Eqs.~(\ref{eq:scattermatch_int}) can be used to calculate the back-scattering coefficients used in Eqs.~(\ref{eq:N-int})\textendash(\ref{eq:N+int}).

\section{\label{app:bog}Evaluation of Bogoliubov coefficient scalar products}

In this Appendix, details are given for the calculation of the inner products of Eqs.~(\ref{eq:bog}) and (\ref{eq:inner_product}) leading to the number operators of Eqs.~(\ref{eq:N}). Focus will be placed on the scalar (spin-0) case, though the final result holds true for any integer-spin modes.

For the scalar product along past null infinity ($\mathscr{I}^-$), where the ingoing Eddington-Finkelstein coordinate $v$ runs from ${-\infty}$ to $\infty$, one may choose
\begin{equation}
    d\Sigma\ n^\mu\sqrt{-g_\Sigma}\ \partial_\mu=dvd(\cos\theta)d\varphi R^2\partial_v,
\end{equation}
while for the scalar product along the past horizon ($\cal{H}_\text{past}={\cal{H}_\text{past}^\text{ext}\cup\cal{H}_\text{past}^\text{int}}$), where the outgoing Kruskal-Szekeres coordinate $U$ runs from ${-\infty}$ to 0 in the interior and from $0$ to $\infty$ in the exterior, one has
\begin{equation}
    d\Sigma\ n^\mu\sqrt{-g_\Sigma}\ \partial_\mu=dUd(\cos\theta)d\varphi_+R_+^2\partial_U.
\end{equation}

The modes to be evaluated along these null hypersurfaces are those of the emitter:
\begin{equation}
    \phi_\text{em}\to
    \begin{cases}
        \displaystyle\frac{\text{e}^{i\bar{m}\varphi}S^{\bar{\omega}}_{\bar{\ell}\bar{m}}(\theta)}{\sqrt{4\pi\bar{\omega}}R}\ \text{e}^{-i\bar{\omega}v},&\mathscr{I}^-\\\\
        \displaystyle\frac{\text{e}^{i\bar{m}\varphi_+}S^{\bar{\omega}}_{\bar{\ell}\bar{m}}(\theta)}{\sqrt{4\pi\bar{\omega}}R_+}\ \text{e}^{-i\bar{\omega}U},&\cal{H}_\text{past}
    \end{cases},
\end{equation}
and those of one of four classes of observers, initialized either at infinity, at the event horizon, or at the left or right leg of the inner horizon:
\begin{equation}
    \phi_\text{ob}\to
    \begin{cases}
        \displaystyle\hat{\mathcal{S}}_v\frac{\text{e}^{\pm im\varphi}S^{\omega}_{\ell m}(\theta)}{\sqrt{4\pi\omega}R}\ \text{e}^{\mp i\hat{\omega}v},&\mathscr{I}^-\\\\
        \displaystyle\hat{\mathcal{S}}_{u_\text{ext}}\frac{\text{e}^{\pm im\varphi_+}S^{\omega}_{\ell m}(\theta)}{\sqrt{4\pi\omega}R_+}\ \text{e}^{\mp i(\hat{\omega}-m\Omega_+)u},&\cal{H}_\text{past}^\text{ext}\\\\
        \displaystyle\hat{\mathcal{S}}_{u_\text{int}}\frac{\text{e}^{\pm im\varphi_+}S^{\omega}_{\ell m}(\theta)}{\sqrt{4\pi\omega}R_+}\ \text{e}^{\mp i(\hat{\omega}-m\Omega_+)u},&\cal{H}_\text{past}^\text{int}
    \end{cases},
\end{equation}
where the upper sign corresponds to the observer modes ${}^\text{ext}\phi_\text{ob}^+$, ${}^\text{ext}\phi_\text{ob}^-$, and ${}^\text{int}\phi_\text{ob}^-$, the lower sign corresponds to the observer modes ${}^\text{int}\phi_\text{ob}^+$, and where quantities with hats also take on different forms for each family of observers:
\begin{align}\label{eq:hat_cases}
    {}^\text{ext}\phi_{\text{ob},\omega}^+&: & \hat{\omega}&=\omega, & &\hat{\mathcal{S}}_v=\mathcal{R}^+_{\text{ext},\omega},\nonumber\\
    & & \hat{\mathcal{S}}_{u_\text{ext}}&=\mathcal{T}^+_{\text{ext},\omega}, & &\hat{\mathcal{S}}_{u_\text{int}}=0,\nonumber\\
    {}^\text{ext}\phi_{\text{ob},\omega}^-&: & \hat{\omega}&=\omega+m\Omega_+, & &\hat{\mathcal{S}}_v=\mathcal{T}^-_{\text{ext},\omega},\nonumber\\
    & & \hat{\mathcal{S}}_{u_\text{ext}}&=\mathcal{R}^-_{\text{ext},\omega}, & &\hat{\mathcal{S}}_{u_\text{int}}=0,\nonumber\\
    {}^\text{int}\phi_{\text{ob},\omega}^-&: & \hat{\omega}&=\omega+m\Omega_-, & &\hat{\mathcal{S}}_v=\mathcal{T}^-_{\text{int},\omega}\mathcal{T}^-_{\text{ext},\omega},\nonumber\\
    & & \hat{\mathcal{S}}_{u_\text{ext}}&=\mathcal{T}^-_{\text{int},\omega}\mathcal{R}^-_{\text{ext},\omega}, & &\hat{\mathcal{S}}_{u_\text{int}}=\mathcal{R}^-_{\text{int},\omega},\nonumber\\
    {}^\text{int}\phi_{\text{ob},\omega}^+&: & \hat{\omega}&=\omega+m\Omega_-, & &\hat{\mathcal{S}}_v=\mathcal{R}^+_{\text{int},\omega}\mathcal{T}^-_{\text{ext},\omega},\nonumber\\
    & & \hat{\mathcal{S}}_{u_\text{ext}}&=\mathcal{R}^+_{\text{int},\omega}\mathcal{R}^-_{\text{ext},\omega}, & &\hat{\mathcal{S}}_{u_\text{int}}=\mathcal{T}^+_{\text{int},\omega}.
\end{align}

Due to the orthogonality of the spheroidal harmonics, the angular pieces can be integrated out to yield Kronecker $\delta$ functions between the observer's and emitter's mode numbers $\ell$ and $m$. Then one has (where the $\pm$ sign is once again defined as above)
\begin{align}\label{eq:scalarproductintegrals}
    \langle\phi_\text{em}|\hat{\phi}_\text{ob}^\text{*}\rangle&=\frac{\delta_{\bar{\ell}\ell}\delta_{(\pm\bar{m})m}}{4\pi i\sqrt{\omega\bar{\omega}}}\bigg(\hat{\mathcal{S}}_v\int_{-\infty}^{\infty}dv\ \text{e}^{-i\bar{\omega}v}\overset{\leftrightarrow}{\partial}_v\text{e}^{\mp i\hat{\omega}v}\nonumber\\
    &+\hat{\mathcal{S}}_{u_\text{ext}}\int_{-\infty}^0dU\ \text{e}^{-i\bar{\omega}U}\overset{\leftrightarrow}{\partial}_U\text{e}^{\mp i(\hat{\omega}-m\Omega_+)u}\nonumber\\
    &+\hat{\mathcal{S}}_{u_\text{int}}\int_0^{\infty}dU\ \text{e}^{-i\bar{\omega}U}\overset{\leftrightarrow}{\partial}_U\text{e}^{\mp i(\hat{\omega}-m\Omega_+)u}\bigg).
\end{align}
In evaluating the bidirectional derivative defined by ${\psi\overset{\leftrightarrow}{\partial}_\mu\phi\equiv\psi\partial_\mu\phi-\phi\partial_\mu\psi}$, the terms in the second and third lines of Eq.~(\ref{eq:scalarproductintegrals}) that have the form ${\partial_U\text{e}^{\mp i(\hat{\omega}-m\Omega_+)u}}$ can be simplified through integration by parts, yielding a surface term that can be safely discarded:
\begin{align}\label{eq:scalarproductintegrals2}
    \langle\phi_\text{em}|\hat{\phi}_\text{ob}^\text{*}\rangle&=\frac{\delta_{\bar{\ell}\ell}\delta_{(\pm\bar{m})m}}{4\pi\sqrt{\omega\bar{\omega}}}\bigg((\bar{\omega}\mp\hat{\omega})\hat{\mathcal{S}}_v\int_{-\infty}^{\infty}dv\ \text{e}^{-i(\bar{\omega}\pm\hat{\omega})v}\nonumber\\
    &+2\bar{\omega}\hat{\mathcal{S}}_{u_\text{ext}}\int_{-\infty}^0 dU\ \text{e}^{-i(\bar{\omega}U\pm(\hat{\omega}-m\Omega_+)u)}\nonumber\\
    &+2\bar{\omega}\hat{\mathcal{S}}_{u_\text{int}}\int_0^\infty dU\ \text{e}^{-i(\bar{\omega}U\pm(\hat{\omega}-m\Omega_+)u)}\bigg).
\end{align}
The observed number operator ${\langle\hat{N}\rangle_{\omega\ell m}}$ can now be evaluated via Eq.~(\ref{eq:bog}) by taking the square of the complex conjugate of Eq.~(\ref{eq:scalarproductintegrals2}) and summing over all emitter modes. The integral in the first line of Eq.~(\ref{eq:scalarproductintegrals2}) reduces to a Dirac $\delta$ function that either vanishes (upper sign) or leaves a small additive factor (lower sign)\textemdash these values can be ascertained by noting that the emitter's modes $\phi_\text{em}$ are normalized along past null infinity as
\begin{align}
    &\langle\phi_\text{em}^{\omega\ell m}|\phi_\text{em}^{\bar{\omega}\bar{\ell}\bar{m}}\rangle=-\langle\phi_\text{em}^{\omega\ell m\text{*}}|\phi_\text{em}^{\bar{\omega}\bar{\ell}\bar{m}\text{*}}\rangle=\delta(\omega-\bar{\omega})\delta_{\ell\bar{\ell}}\delta_{m\bar{m}},\nonumber\\
    &\langle\phi_\text{em}^{\omega\ell m}|\phi_\text{em}^{\bar{\omega}\bar{\ell}\bar{m}\text{*}}\rangle=0.
\end{align}
The integrals in the second and third lines of Eq.~(\ref{eq:scalarproductintegrals2}), on the other hand, are the origin of the Planckian distribution. Using the definition of $U$ from Eq.~(\ref{eq:U}), these can be evaluated in terms of $\Gamma$ functions \cite{bar11a}. The resulting number operator can then be found after taking the modulus squared:
\begin{align}\label{eq:scalarproductintegrals3}
    &\langle\hat{N}\rangle_{\omega\ell\{\pm m\}}=\frac{1}{4\pi^2\omega}\int_0^\infty d\bar{\omega}\ \bar{\omega}\ \bigg|\left\{\genfrac{}{}{0pt}{}{0}{2\pi\hat{\mathcal{S}}_v\delta(\bar{\omega}-\hat{\omega})}\right\}\nonumber\\
    &+\left(\hat{\mathcal{S}}_{u_\text{ext}}+(-1)^{-z}\hat{\mathcal{S}}_{u_\text{int}}\right)\Gamma(z)(-i\bar{\omega})^{-z}\varkappa_+^{z-1}\bigg|^2,
\end{align}
where we have defined the quantity
\begin{equation}
    z\equiv1\pm\frac{i(\hat{\omega}-m\Omega_+)}{\varkappa_+}.
\end{equation}
The squared modulus of Eq.~(\ref{eq:scalarproductintegrals3}) can be simplified by the property
\begin{equation}
    \left|\Gamma\left(1+bi\right)\right|^2=\frac{\pi b}{\sinh(\pi b)}.
\end{equation}
First, consider the upper sign of Eq.~(\ref{eq:scalarproductintegrals3}), corresponding to the observer modes ${}^\text{ext}\phi_\text{ob}^+$, ${}^\text{ext}\phi_\text{ob}^-$, and ${}^\text{int}\phi_\text{ob}^-$ (i.e.,\ all modes except those originating from an outgoing observer at the inner horizon, who has exp(${i\omega t}$) instead of exp(${-i\omega t}$)). Then, assuming the frequencies $\hat{\omega}$ and $\bar{\omega}$ are strictly positive, the first term in the integrand of Eq.~(\ref{eq:scalarproductintegrals3}) will vanish, and the observed number operator will simplify to
\begin{align}\label{eq:scalarproductintegrals4}
    \langle\hat{N}\rangle_{\omega\ell m}&=\frac{\hat{\omega}-m\Omega_+}{\omega}\frac{\left|\hat{\mathcal{S}}_{u_\text{ext}}-\text{e}^{\frac{\pi}{\varkappa_+}(\hat{\omega}-m\Omega_+)}\hat{\mathcal{S}}_{u_\text{int}}\right|^2}{\text{e}^{\frac{2\pi}{\varkappa_+}(\hat{\omega}-m\Omega_+)}-1}\nonumber\\
    &\times\frac{1}{2\pi\varkappa_+}\int_0^\infty\frac{d\bar{\omega}}{\bar{\omega}}.
\end{align}
The integral over the emitter's frequency modes in the second line of Eq.~(\ref{eq:scalarproductintegrals4}) diverges, but such behavior is not a problem and is actually to be expected. The divergence originates from the use of continuum-normalized plane waves to initialize the modes detected over an entire Cauchy hypersurface, which inevitably leads to an infinite amount of Hawking radiation reaching an observer throughout the infinite amount of time left in the future. If instead of plane waves, a more physically realistic choice is used to represent the observer's modes, such as a finite wave packet distribution, then the resulting integral will be regularized and the second line in Eq.~(\ref{eq:scalarproductintegrals4}) will reduce to unity \cite{fab05}. In particular, one may model the observer as a particle detector sensitive only to frequencies within a small $\epsilon$ of ${\omega\sim j\epsilon\sim(j+1)\epsilon}$, which is turned on at a time ${u=2\pi n/\epsilon}$ for a duration ${2\pi/\epsilon}$ (for integers $j$ and $n$). Then the observer's modes will appear as
\begin{equation}\label{eq:wavepacket}
    \phi_\text{ob}^\text{reg}=\frac{1}{\sqrt{\epsilon}}\int_{j\epsilon}^{(j+1)\epsilon}d\omega\ \text{e}^{2\pi i\omega n/\epsilon}\phi_\text{ob}.
\end{equation}
At late times (large $n$), the above expression will yield exactly the first line of Eq.~(\ref{eq:scalarproductintegrals4}), with the remaining terms on the second line reducing to unity. This expression for the expectation value of the number operator ${\langle\hat{N}\rangle_{\omega\ell m}}$ reproduces Eqs.~(\ref{eq:N+ext})\textendash(\ref{eq:N-int}) after substituting the respective observer quantities from Eqs.~(\ref{eq:hat_cases}).

Finally, consider the lower sign of Eq.~(\ref{eq:scalarproductintegrals3}), corresponding to the observer modes ${{}^\text{int}\phi_\text{ob}^+}$. As in the previous case, it will be helpful to consider a wave-packet version of the observer's modes, since the integral over the complex square modulus of a Dirac delta distribution must be regularized in some meaningful way.

For the portion of the observer's modes along past null infinity, performing the integral of Eq.~(\ref{eq:wavepacket}) gives
\begin{equation}
    \phi_\text{ob}^\text{reg}\propto\hat{\mathcal{S}}_v\frac{\sqrt{\epsilon}}{\frac{v\epsilon}{2}+n\pi}\ \text{e}^{i\frac{v\epsilon}{2}(2j-1)}\sin\left(\frac{v\epsilon}{2}\right).
\end{equation}
The Fourier transform of these modes over the emitter's frequency yields the inner product
\begin{equation}
    \langle\phi_\text{em}|\hat{\phi}_\text{ob}^{\text{reg*}}\rangle_{\mathscr{I}^-}=\delta_{\bar{\ell}\ell}\delta_{(-\bar{m})m}\hat{\mathcal{S}}_v\sqrt{\frac{\bar{\omega}}{\omega}}\ \frac{\text{e}^{2\pi i\bar{\omega}n/\epsilon}}{\sqrt{\epsilon}}
\end{equation}
in the frequency range ${(j-1)\epsilon<\bar{\omega}<j\epsilon}$, and 0 everywhere else. Once this quantity's complex modulus is squared and summed over the emitter's modes, a constant term will remain of the form
\begin{equation}
    \frac{(2j-1)\epsilon}{2\omega}\hat{\mathcal{S}}_v\sim\frac{\hat{\omega}}{\omega}\hat{\mathcal{S}}_v,
\end{equation}
since for small $\epsilon$ the quantity $j\epsilon$ is precisely the observer's frequency ${\hat{\omega}}$ back-propagated to past null infinity. Thus, the number operator of Eq.~(\ref{eq:bog}) for an outgoing observer at the inner horizon reduces to
\begin{align}\label{eq:scalarproductintegrals5}
    &\langle\hat{N}\rangle_{\omega\ell(-m)}=\nonumber\\
    &-\frac{\hat{\omega}-m\Omega_+}{\omega}\frac{\left|\hat{\mathcal{S}}_{u_\text{ext}}-\text{e}^{-\frac{\pi}{\varkappa_+}(\hat{\omega}-m\Omega_+)}\hat{\mathcal{S}}_{u_\text{int}}\right|^2}{\text{e}^{-\frac{2\pi}{\varkappa_+}(\hat{\omega}-m\Omega_+)}-1}+\frac{\hat{\omega}}{\omega}\hat{\mathcal{S}}_v.
\end{align}
Note that in principle Eq.~(\ref{eq:scalarproductintegrals5}) will contain an additional cross term when the complex modulus of Eq.~(\ref{eq:scalarproductintegrals3}) is squared, which can be expressed in terms of incomplete gamma functions. However, in the late-time limit of large $n$, this term and its complex conjugate become negligibly small and thus are not included here.

The expression in Eq.~(\ref{eq:scalarproductintegrals5}) for the expectation value of the number operator ${\langle\hat{N}\rangle_{\omega\ell m}}$ reproduces Eq.~(\ref{eq:N+int}) after substituting the respective observer quantities from Eqs.~(\ref{eq:hat_cases}).

\bibliography{apsbib}

\end{document}